%% file: main.tex
\pdfoutput=1
\documentclass[12pt,onecolumn]{IEEEtran}
\usepackage{amsmath,amsfonts}
\usepackage{algorithmic}
\usepackage{array}
\usepackage[caption=false,font=normalsize,labelfont=sf,textfont=sf]{subfig}
\usepackage{textcomp}
\usepackage{stfloats}
\usepackage{url}
\usepackage{verbatim}
\usepackage{graphicx}
\usepackage{xcolor}
\usepackage[acronyms,nonumberlist,nopostdot,nomain,nogroupskip]{glossaries}
\input{acronyms.tex}
\usepackage{booktabs}
\usepackage{siunitx}
\usepackage{svg}
\usepackage{changepage}

\usepackage{setspace}   

\def\BibTeX{{\rm B\kern-.05em{\sc i\kern-.025em b}\kern-.08em
    T\kern-.1667em\lower.7ex\hbox{E}\kern-.125emX}}
\usepackage{balance}
\begin{document}
\title{Sea Trial Validation of the ROS-DESERT Middleware with Autonomous Underwater Vehicles}
\author{
Davide Cosimo\textsuperscript{1,2}, Davide Costa\textsuperscript{3}, Riccardo Costanzi\textsuperscript{1}, 
Filippo Campagnaro\textsuperscript{3}, \\
Andrea Caiti\textsuperscript{1}, Michele Zorzi\textsuperscript{3} \\[1em] 

\textsuperscript{1}Dept. of Information Engineering, University of Pisa, Italy \\
\textsuperscript{2}Naval Support and Experimentation Centre - Italian Navy, La Spezia, Italy \\
\textsuperscript{3}Dept. of Information Engineering, University of Padua, Italy \\[1em] 

\texttt{davide.cosimo@phd.unipi.it, costadavid@dei.unipd.it, riccardo.costanzi@unipi.it,}\\
\texttt{filippo.campagnaro@unipd.it, andrea.caiti@unipi.it, michele.zorzi@unipd.it}

\thanks{This manuscript has been submitted for consideration for publication in the IEEE Journal of Oceanic Engineering (IEEE JOE).}
}

\markboth{}%
{}

\maketitle

\doublespacing        %
\pagestyle{plain} 

\begin{abstract}
This paper presents a modular software architecture that enables environmental-aware coordination of heterogeneous~\glspl{auv} to improve underwater acoustic connectivity. The architecture combines a  Robot Operating System 2 application layer with the DESERT Underwater communication framework through the \emph{rmw\_desert} middleware, and integrates a Robot Operating System 1 bridge to ensure interoperability with legacy vehicle front-seat controllers. This design enables fine-grained, cross-layer configurability of the communication stack and supports onboard processing of environmental measurements to inform adaptive communication behaviors.

As a representative use case, this architecture is used to implement a lightweight depth-optimization strategy that exploits environmental awareness and \gls{auv} mobility to improve acoustic link performance. The complete software stack is validated through sea trials conducted off the Gulf of La Spezia in littoral water with an average depth of approximately 100\,m using a deployment involving three \glspl{auv} with distinct operational roles. Experimental results indicate that depth-adaptive repositioning yields measurable gains in packet reception at horizontal separation of approximately 1\,km, while differences are negligible at shorter ranges where the received signal energy remains above demodulation thresholds. Beyond link-level performance the sea trials confirm the feasibility, modularity, and practical deployability of the proposed architecture on existing~\gls{auv} platforms. 
\end{abstract}

\begin{IEEEkeywords}
\noindent AUVs, Underwater acoustic communication, Depth optimization, DESERT Underwater, ROS interoperability
\end{IEEEkeywords}

\glsresetall

\section{Introduction}
The deployment of~\glspl{auv} has grown rapidly in recent years, driven by their lower per-mission cost, operational flexibility, and ability to access hazardous or otherwise inaccessible environments. Compared to legacy crewed methods,~\glspl{auv} reduce logistical burden, increase mission persistence, and enable fine-grained adaptive sampling across a wide range of tasks. Modern systems such as the REMUS 6000 and Hugin~\cite{WHOI_Dalio2019_remus6000, Zwolak_2020_hugin} exemplify the diversity of platforms now routinely employed in ocean science, offshore industry, and defense operations. This versatility is accelerating the adoption of~\glspl{auv} across scientific, commercial, and defense applications, and is motivating ever larger and more heterogeneous multi-vehicle deployments. A fundamental enabler and frequent bottleneck for these deployments is underwater communication. Although alternative modalities (e.g., optical or magneto-inductive links~\cite{SummarySensorNet}) have proven useful in short-range deployments, acoustic communication remains the only practical and broadly applicable solution for long-range transmissions~\cite{Lurton, Stojanovic2002_HandbookChapter}. However, acoustic channels are severely constrained by low bandwidth and long, variable propagation delays, as well as by environmental effects that induce multipath and spatial, temporal, and frequency variability. These physical limitations profoundly influence the design of multi-\gls{auv} coordination protocols and distributed sensing algorithms~\cite{Caiti_swarm}.


The main contribution of this paper is the experimental validation and system-level assessment of the open-source framework presented in~\cite{RmwDesert}, developed by the Department of Information Engineering at the University of Padova to enable efficient data transmission from~\gls{ros2} to the underwater environment. While~\cite{RmwDesert} described the design of the \emph{rmw\_desert} middleware and its integration with the~\gls{desert} Underwater communication stack, this work extends that contribution by demonstrating the complete architecture in real-world sea trials and evaluating its performance in an operational multi-\gls{auv} scenario. In addition, we provide a more detailed system-level description of the architecture.
Unlike many existing approaches that focus on specific algorithms or hardware, the proposed architecture integrates several key components to address the challenges of underwater communication. It combines high data compression techniques with the~\gls{desert} Underwater protocol stack~\cite{Desert, Desert2}, a proven solution for reliable communication in underwater settings. One of the key features of this system is its flexibility: it can be used in both simulated environments and real-world sea trials, allowing researchers to test and refine their communication strategies in various stages of development. This capability makes the framework highly adaptable to different types of research projects, from early-stage simulations to final deployment in the field. Another major advantage is that the framework does not depend on a specific brand or model of acoustic modem. Instead, it supports a wide range of modems, including those from vendors and research institutes such as Evologics, SuM, and Ahoi, making it more accessible to researchers working with different hardware. Moreover, the system can be easily extended to support other modem types by adding a small software module, allowing future-proofing and integration with new technologies as they emerge.

The system's architecture is designed to handle the entire communication stack, from the physical layer to the application layer, while presenting a simple and clear interface for users. This design allows users to focus on defining the behavior and mission objectives of~\glspl{auv} without having to manage the complexities of the underlying communication processes. To demonstrate its effectiveness, we tested the framework in a use case scenario where mobility and environmental measurements were used to dynamically adjust node positions. This adaptation aimed to improve connectivity and communication efficiency in multi-vehicle deployments. By adjusting the relative positions of the vehicles based on real-time data, the system was able to optimize the network performance, ensuring more stable and reliable communication across the deployment area. This approach underscores the framework’s capability to support complex, real-world scenarios where communication conditions vary dynamically.

The paper is structured as follows. Section \ref{stateofart} presents a comprehensive state-of-the-art analysis. Section \ref{preliminaries} provides the necessary preliminaries, including the software tools and hardware platforms employed in this study. Section \ref{systemdescription} describes the complete system architecture. Section \ref{setupandsettings} details the sea-trial setup and experimental procedures. Section \ref{results} presents the results and quantitative evaluation. 
Finally, Section \ref{conclusions} concludes the paper and outlines directions for future work.

\section{State of the art} \label{stateofart}
This section reviews the state of the art in vehicle control frameworks compatible with underwater networks (Section \ref{ctrl_frameworks}) and in communication optimization strategies employed for these types of networks (Section \ref{comm_optimization}).

\subsection{Control frameworks} \label{ctrl_frameworks}
Controlling vehicles via underwater communication links presents substantial challenges due to the limited bandwidth, long propagation delays, and strong variability of the underwater acoustic channel~\cite{acoustic_comms_advances, acoustic_comms_future}. To address these constraints, a number of software frameworks and communication stacks have been proposed in recent years to support underwater networking, particularly for mobile autonomous platforms such as \glspl{auv}. These solutions differ considerably in architectural scope, protocol-layer coverage, and degree of modularity, which in turn affects their suitability for closed-loop vehicle control applications.

A key distinction among existing approaches lies in the extent of the communication stack that they implement or manage. Some frameworks are restricted to the application layer, where they provide standardized message formats or mission-level communication services while relying on existing transport protocols and lower-layer infrastructure~\cite{catl, moos}. Such lightweight solutions generally facilitate the integration with established \gls{auv} software architectures; however, they depend on external mechanisms to support essential networking functions, including reliable data delivery, medium access control, and routing.

Other approaches adopt a complete design by implementing an end-to-end communication stack integrating the physical, data link, network, transport, and application layers~\cite{guwmanet}. These full-stack frameworks aim to control all aspects of underwater communication, from modulation and synchronization at the physical layer to addressing, routing, and inter-node coordination at higher layers. While this integrated approach can provide improved robustness and adaptability—particularly in dynamic or multi-hop network scenarios—it also introduces increased system complexity and higher computational and energy requirements~\cite{UWchallenge}. Consequently, such solutions may be suitable for smaller \gls{auv} platforms with stringent resource constraints.

The selection between lightweight and full-stack communication architectures is driven by a combination of mission-level and system-level considerations. These include mission characteristics and communication range, as long-range operations may benefit from cross-layer designs in which PHY-aware medium access control and routing strategies are used to mitigate high latency and channel variability~\cite{phy_aware}. Platform resource constraints also play a key role, since vehicles with limited processing capacity, energy availability, or acoustic hardware capabilities may only support partial protocol stack implementations. Interoperability requirements further influence the choice, as application-layer solutions based on standardized interfaces generally simplify the integration across heterogeneous platforms and systems. Finally, scalability and autonomy considerations are important, with full-stack frameworks being more suitable for supporting autonomous behavior in complex or decentralized network topologies, such as mesh and swarm configurations.

This diversity of architectural approaches highlights the absence of a universally optimal solution for wireless underwater remote vehicle control~\cite{ava}. Depending on the operational context, system designers may favor reduced complexity and ease of integration over architectural completeness and adaptability, or vice versa. For example, the evaluation and selection of a communication framework for long-range underwater missions should consider not only the achievable communication range and physical-layer performance, but also the extent of protocol-layer coverage and its alignment with overall system architecture and resource constraints.

Protocols such as \gls{guwmanet}~\cite{guwmanet} enable the formation of robust, delay-tolerant mobile ad hoc underwater wireless networks. When used in conjunction with the \gls{guwal}~\cite{guwal}, the full-stack framework supports efficient multicast routing with minimal overhead by employing a compact, general-purpose application data format tailored to the stringent bandwidth and energy constraints of underwater acoustic communications. The design accommodates diverse communication requirements while maintaining very short packet lengths, thereby reducing protocol overhead and transmission duration. In particular, \gls{guwal} defines a set of parcel types to support different underwater communication functions, including data requests as well as command-and-control messaging.

Another representative approach is the \gls{catl}~\cite{catl}, a lightweight application layer protocol specification intended to support coordination and task allocation among autonomous systems, with particular emphasis on \glspl{auv}. \gls{catl} does not constitute a physical or network-layer communication protocol; rather, it defines a set of structured message formats specified using the \gls{json} and the \gls{dccl}~\cite{dccl}, without providing native mechanisms for data transmission or network-level functionality. These standardized message definitions facilitate semantic interoperability across heterogeneous autonomous platforms, enabling the exchange of mission-relevant information in collaborative operational contexts, while leaving the choice of underlying transport mechanisms to the system designer.

A different strategy can be implemented using ros\_acomms~\cite{ros_acomms}, a \gls{ros} package designed to support the transmission of \gls{ros} messages over acoustic links. The software provides message serialization, compression, and an integrated protocol stack for medium access control and modem interfacing, with native support limited to acoustic modems from the WHOI family. Communication is realized by encapsulating \gls{ros} messages within a fixed application-layer protocol designed to take into account the constraints of acoustic channels, enabling the exchange of mission-level information between underwater robotic platforms. The module is distributed as a standalone package outside the official \gls{ros} repositories and is typically deployed in specific scenarios that do not require network topology or protocols customization.

While these frameworks demonstrate the effectiveness of lightweight, application-oriented solutions for underwater coordination and messaging, they are not general-purpose control architectures. In practice, most of the existing solutions primarily provide communication primitives or standardized message formats, leaving the design and integration of control logic, coordination strategies, and system-level behaviors largely application-specific. As a result, each control application must typically be developed from scratch and tightly tailored to the target mission, platform capabilities, and chosen transport mechanisms, limiting reusability and portability across different scenarios.

In contrast, the approach investigated in this work aims to provide a more general and reusable solution for \gls{auv}-based systems by leveraging the combination between the \gls{ros} and the \gls{desert} Underwater networking framework initially presented in~\cite{RmwDesert}. By embedding underwater acoustic communications directly within the \gls{ros} middleware abstraction, this approach enables control, coordination, and autonomy algorithms to be developed independently of the specific characteristics of the underlying acoustic network. This promotes modularity and code reuse, allowing the same high-level control applications to operate across different missions and network configurations, while relying on \gls{desert} to handle the complexities of underwater communications. As such, the proposed approach moves toward a general control and communication framework rather than a collection of application-specific solutions.

\subsection{Communication optimization} \label{comm_optimization}
Research on~\glspl{uan} has intensified over the last twenty years, motivated by the need to populate the underwater domain with nodes that support scientific exploration, environmental monitoring, offshore industry, and defense applications~\cite{ComprehensiveSummary_2022}.

Despite this strong momentum, the vision of a pervasive~\gls{iout} and the emergence of “big marine data” face substantial technological and operational obstacles~\cite{Jahanbakht2021_IoUT, Qiu2020_UIoT}. These include material degradation under extreme pressure and corrosive seawater conditions~\cite{Liu2022_HydrostaticPressure,Hao2024_degradation}, and the accumulation of biological fouling on exposed surfaces~\cite{Oiler2015_biofouling, Vuong2023_biofouling}; complex and costly logistics for deployment, recovery and maintenance; difficulties in achieving accurate node localization without~\gls{gps}~\cite{Paull2014_AUVNavReview}; and severely constrained energy budgets, since most subsea platforms rely on batteries, in the absence of a seabed power infrastructure~\cite{Islam2022_energyUWC}.

Above all, the intrinsic limitations of underwater communications remain the primary bottleneck for reliable multi-node networking. Underwater communications exploit several physical modalities — acoustic, optical,~\gls{rf} and~\gls{mi} — each with distinct trade-offs. Acoustic links provide the longest practical ranges but suffer from limited bandwidth, long and variable propagation delays, strong multipath and pronounced spatio-temporal variability~\cite{Stojanovic2009_channelModels}; optical links offer high data rates over short ranges and are strongly affected by turbidity, external light interference, and alignment~\cite{Sun2020_UWOC,OpticExp_Caiti};~\gls{rf} suffers extreme attenuation in water and is typically limited to very short ranges or shallow, low-salinity deployments;~\gls{mi} enables robust very-short-range links (e.g., through seawater or pipes) but at low data rates~\cite{Li2019_MagInduction}. These modality-specific constraints explain why acoustic communication remains the dominant choice for long-range underwater networking, despite its severe limitations.~\cite{Stojanovic2002_HandbookChapter}

Foundational and more recent surveys in~\gls{uan} provide a comprehensive overview of channel impairments, protocol limitations, and architectural trade-offs that must be addressed in system development~\cite{UW_comms_tec_review, UWSN_survey_2019, Aman2022_UnderwaterAirWaterSurvey}.
Research on mobility-assisted networking has produced several families of techniques that exploit node movement as an active degree of freedom to improve connectivity, throughput and energy efficiency in underwater networks. For example, the UAN11 field experiment in Norwegian waters deployed a heterogeneous system composed of a ~\gls{stu}, three~\glspl{fno} and mobile Fòlaga~\glspl{auv}~\cite{Alvarez2009_Folaga,Caffaz2010_Folaga} to evaluate topology-adaptation strategies in realistic conditions \cite{Caiti_UAN11}. The campaign demonstrated practical mobility-driven behaviors: vehicles autonomously detected link loss and re-positioned to restore acoustic coverage, thereby validating the use of on-board autonomy and topology reconfiguration to optimize network reachability and resilience in the field.  
Following this practical validation, the literature presents several algorithmic paradigms: optimization-based node selection, probabilistic topology-control, and mobility-aware geographic routing that proactively steer nodes to avoid voids and shadow zones. These approaches trade navigation cost (energy, time) for improved network metrics (connectivity, latency, delivery probability) and have been mainly investigated through theoretical analysis and simulation studies~\cite{Coutinho2018_mobilitytoimprove, Coutinho2013_movassistedtopology,Li2017_probabilistic}.

A second group leverages mobility for relay/ferrying and data muling: mobile~\glspl{auv} act as deliberate relays or data ferries between otherwise disconnected clusters or between seabed sensors and surface gateways. Such techniques are particularly attractive when a persistent infrastructure is unavailable; they range from simple store-and-forward policies to trajectory-optimized relaying where path planning is co-designed with communication objectives (e.g., maximizing delivered data or minimizing latency). Recent works demonstrate~\gls{auv}-aided isolated-subnetwork prevention and~\gls{auv}-enabled offloading frameworks that integrate mobility, scheduling and link-layer parameters~\cite{AUVaided_2024}.

Early studies recognized the potential of vehicle mobility to mitigate connectivity gaps in underwater networks. For example, the study in~\cite{MultipleUUV_2005} proposes using multiple~\glspl{uuv} to patrol regions where temporal interference or poor initial deployment creates connectivity gaps. In the proposed scheme,~\glspl{uuv} identify critical communication holes between underwater sensors and act as mobile bridges to restore links; when needed, they can deploy additional sensors in physically disconnected areas or serve as temporary local sinks for isolated partitions, delivering collected data to the nearest connected segment of the network. While this approach demonstrated the potential of mobility to restore network connectivity, it also highlighted the significant energy costs associated with continuous vehicle motion.

More recent work has investigated data-muling and cluster-based strategies that explicitly trade mobility for communication efficiency. In the context of the~\gls{robovaas} project, the study in~\cite{Signori2019_robovaas} proposed a polling-based~\gls{mac} protocol (UW-POLLING) for underwater data collection scenarios, where a mobile cluster-head~\gls{auv} collects data from static sensor nodes and optionally forwards them to a shore-connected sink through above-water links.

An additional example is~\cite{PAMP_coPAMP_2017}, which applies~\gls{pamp}/~\gls{co-pamp} with gliders that occupy predefined sojourn positions and adapt motion and transmit power to improve reachability. Simulations report an increased packet delivery ratio, fewer void zones, and reduced energy consumption, highlighting the energy–connectivity trade-off that motivates energy-aware mobility designs.

From a communications perspective, mobility is frequently combined with adaptive physical- and~\gls{mac}-layer techniques. Recent literature proposes adaptive~\gls{mac} protocols specifically designed for mobile~\gls{auv} formations and mobile~\glspl{uasn}: a scheduling-based~\gls{abmac} uses an incompletely centralized frame scheduler (a control node per frame), variable slot lengths and adaptive transmission timing (optimized via a genetic algorithm) to avoid collisions and minimize the frame length, achieving improved throughput and robustness even in poor channels; conversely, a load-adaptive~\gls{csmaca} scheme (LACC-M) provides two transmission modes and a special broadcast join packet so that mobile nodes can dynamically switch modes according to the network load, yielding higher throughput, lower delay, and better energy efficiency than prior contention-based protocols. These adaptive~\gls{mac} designs—when combined with mobility and cross-layer adaptation of PHY/\gls{mac} parameters—have been shown (in simulation) to substantially enhance timeliness, reliability and energy performance in realistic mobile underwater scenarios~\cite{AdaptiveMAC_2023,AdaptiveCSMA_2018}.

In~\cite{ModelBased_2013} the authors propose a hybrid model and data driven framework that enables~\glspl{auv} to adapt their depth in real time to maximize acoustic connectivity and sensing performance. The approach couples the~\gls{moosipv} autonomy architecture~\cite{MOOS_2010} with the~\gls{gram}, an embedded acoustic modeling service that runs physics-based propagation models (e.g., Bellhop ~\cite{Bellhop_1992,Bellhop_manual}, Kraken~\cite{Kraken_2001}) onboard or on a low-power model farm to forecast transmission loss and array/~\gls{snr} metrics. The autonomy layer converts these forecasts into depth-oriented utility functions and executes behaviors accordingly. 

Building on these insights, our use case scenario investigates a similar depth-optimization strategy at a swarm level, with a focus on collective communication performance~\cite{Cosimo_Oceans2025}. In our approach, only a designated leader~\gls{auv} is equipped with a ~\gls{ctd} probe and computes the~\gls{ssp}, from which it derives the optimal communication depth, typically near the~\gls{sofar} duct. This information is then disseminated to the other swarm members, which adjust their depth accordingly. We implemented and evaluated this strategy using the~\gls{desert} Underwater Simulator~\cite{Desert,Desert2}, integrated with the~\gls{woss} for realistic acoustic propagation modeling~\cite{WOSS_Casari,opensourceDESERTandWOSS2014}. Different sensing policies were tested, ranging from a single~\gls{ssp} measurement at mission start to multiple updates during the mission. Simulation scenarios considered swarms of 3, 5 and 7 nodes operating under both~\gls{csma} and~\gls{tdma}~\gls{mac} schemes. Results show that depth optimization consistently reduces~\gls{per} and improves throughput, with performance gains more pronounced when periodic~\gls{ctd} updates are available.

\section{Preliminaries} \label{preliminaries}

This section provides the foundational background necessary for the experimental work. It introduces \gls{ros2} (Section \ref{ros_intro}), the~\gls{desert} Underwater framework (Section \ref{desert_intro}), and the X300~\gls{auv} platform used in the trials (Section \ref{x300_intro}).

\subsection{Introduction to ROS2} \label{ros_intro}
The~\gls{ros2} framework is a modern, open-source set of libraries designed to support the development of modular, distributed, and robust robotic systems used as the de-facto standard in industry~\cite{RosStudy}. Building on the lessons learned from~\gls{ros1}, its second major release introduces a fully restructured architecture composed of clearly defined layers that decouple the application logic from system-level communication and transport. This layered design enhances maintainability, scalability, and real-time performance, key requirements for robotics applications operating in complex and dynamic oceanic environments.

At the core of this architecture is the middleware layer, which serves as the communication backbone of~\gls{ros2}~\cite{RosDesign}. Unlike~\gls{ros1}, which relied on a custom transport protocol with a centralized leader node,~\gls{ros2} adopts the Data Distribution Service (DDS) as its default middleware interface. DDS is an open standard for data-centric publish/subscribe communication, designed for real-time, high-performance, and fault-tolerant systems. This change enables decentralized communication between~\gls{ros2} nodes, eliminating the single point of failure and making the system more resilient—an essential feature for marine systems such as~\glspl{auv}, remotely operated vehicles (ROVs), and sensor networks deployed in unpredictable underwater conditions.

The middleware layer is abstracted via the~\gls{ros2} middleware interface, which allows developers to choose between different implementations such as Fast DDS, Cyclone DDS, and RTI Connext, depending on system requirements, such as latency, determinism, or resource usage without changing the core user application logic each time~\cite{RosDesign}. This abstraction provides flexibility to tailor deployments for diverse engineering applications, from low-power embedded nodes on buoys to high-bandwidth surface vessels coordinating a fleet of subsea robots.

Above the middleware layer,~\gls{ros2} features a runtime system responsible for node lifecycle management, intra-process communication, and tools for scheduling and execution. The top application layer offers a rich ecosystem of reusable packages for perception, control, navigation, and simulation. Together, these layers form a cohesive framework that supports the design of complex robotic systems capable of operating autonomously and reliably in real-world marine environments.

The layered architecture of~\gls{ros2} and, in particular, its robust middleware foundation, position it as a powerful tool for ocean engineers and researchers seeking to build scalable, interoperable and fault-tolerant robotic systems for underwater exploration, environmental monitoring and offshore infrastructure inspection.

\subsection{Introduction to DESERT Underwater} \label{desert_intro}
The proliferation of marine applications such as environmental monitoring, offshore exploration, underwater surveillance, and autonomous ocean sampling has driven significant advancements in underwater communication systems. Among the open-source frameworks built to support these developments,~\gls{desert} Underwater~\cite{Desert},~\cite{Desert2} has emerged as a prominent and widely adopted software suite. Built on top of the well-known Network Simulator 2, this set of libraries provides a modular and extensible environment for testing underwater acoustic networks.

Developed by the SIGNET group in the Department of Information Engineering at the University of Padova,~\gls{desert} Underwater was designed to fill a critical gap in the community: the lack of a flexible, standard, and reproducible platform for the performance evaluation of underwater network protocols. Traditional network simulators often fell short in accurately modeling the physical and MAC layers of underwater communications, especially in capturing the unique characteristics of the underwater acoustic channel—such as low bandwidth, high latency, Doppler effects, and signal attenuation.~\gls{desert} addresses these challenges by integrating accurate acoustic propagation models, customizable protocol stacks, and interfaces for real-time testing and deployment.

One of the defining features of~\gls{desert} Underwater is its layered architecture, which allows researchers to plug in custom modules at the physical,~\gls{mac}, network, transport, and application layers. This modularity makes it ideal for testing new algorithms and performing comparative analysis. The framework also includes tools for both simulation and emulation, enabling a transparent transition from theoretical design to real-world implementation. Through its emulation capabilities,~\gls{desert} can interface with actual acoustic modems, enabling hardware testing and real-time experimentation.

A key component of the architecture used in this work is the \verb+uwApplication+ layer, which represents the highest level of the~\gls{desert} protocol stack. It is responsible for handling data packets from user programs according to an application-specific logic. The \verb+uwApplication+ layer enables the implementation of real-world behaviors such as periodic data collection, event-driven communication, remote control signaling, and~\gls{auv} mission coordination. By abstracting the application behavior from the lower layers, it provides flexibility for researchers and developers to design and test realistic use cases.

Beyond its modularity and extensibility, one of the most powerful aspects of~\gls{desert} Underwater lies in its ability to manage and customize the entire communication protocol stack, fully adhering to the ISO/OSI reference model. Each layer of the stack, from the physical up to the application layer, can be independently implemented, modified, or replaced, allowing researchers to experiment with innovative communication strategies or optimize existing ones for specific underwater conditions. This high degree of configurability not only enables a fine analysis of the interactions among protocol layers but also facilitates cross-layer optimization, a key aspect in underwater acoustic networks where physical constraints heavily influence higher-level behaviors.

Furthermore,~\gls{desert} extends beyond pure simulation: through its emulation framework, it can directly interface with commercial acoustic modems, effectively acting as a driver for the specific hardware in use. This feature bridges the gap between theoretical design and practical deployment, providing an environment where approaches tested in simulation can be validated in real time with physical devices. Such integration significantly enhances reproducibility and accelerates the transition from research to field experimentation, making~\gls{desert} Underwater an important tool for both academic and industrial applications.

Over the years,~\gls{desert} was employed in numerous scenarios and has supported a wide range of academic and industrial research initiatives, including projects like the CommsNet’13 experience, where the framework was used in the context of a collaboration with the NATO STO Centre for Maritime Research and Experimentation to measure various network statistics during a remote data retrieval in underwater networks~\cite{commsnet}. 
More recently,~\gls{desert} has been adopted in large-scale integration efforts such as the~\gls{robovaas} project, where it contributed to the development and validation of a hybrid underwater/above-water communication infrastructure for port services and underwater sensor data collection~\cite{Coccolo2023_robovaas}. These deployments further demonstrate the maturity of~\gls{desert} as a tool not only for protocol-level studies, but also for the development and validation of end-to-end underwater communication systems. Its compliance with standardization efforts in underwater networking also makes it a relevant tool for organizations aiming to develop communication protocols that align with international benchmarks.

\subsection{X300 Autonomous Underwater Vehicles} \label{x300_intro}
The X300 is a torpedo-shaped~\gls{auv} developed by the Italian company GraalTech~\cite{GraalTech}, based in Genoa, as the evolution of the previous Folaga model~\cite{Alvarez2009_Folaga, Caffaz2010_Folaga}. The main features of the standard X300 configuration are summarized in Table~\ref{tab:X300}. One of its key advantages is its modular design, which allows the integration of a wide range of payloads in the central section. 

\begin{table}[h]
    \centering
    \small
    \renewcommand{\arraystretch}{1.3}
    \caption{X300 main features}
    \begin{tabular}{p{4.5cm} p{10cm}}
        \toprule
        \textbf{Parameter} & \textbf{Value} \\
        \midrule
        Length & 2220 mm \\
        Weight in air & 29 kg \\
        Weight in water & \SIrange{-0.35}{0.35}{\kilogram} \\ 
        Diameter & 155 mm \\
        Maximum depth & 300 m \\
        Maximum speed & 5 kn \\
        Endurance & 14 h at maximum speed \\
        Batteries & Li-Ion 24 V -- 1200 Wh \\
        Navigation sensors & GPS, depth gauge, compass, 3D tilt sensor \\
        Further sensors & Humidity, Pressure, Battery charge sensor \\
        \bottomrule
    \end{tabular}
    \label{tab:X300}
\end{table}

As in many contemporary~\gls{auv} architectures~\cite{Backseat},~\cite{AutonomyVSComms}, the X300 preserves the well-established \textit{front-seat / back-seat} separation. In the X300 the front-seat subsystem is dedicated to low-level sensing, real-time navigation, attitude and depth control, and the supervision of vehicle vitals; the back-seat hosts mission planning, payload-specific processing and higher-level autonomy, issuing setpoints to the front-seat controller as required. The X300 is designed so that the back-seat can run standard software such as the~\gls{ros}~\cite{ROSOrg}, while the front-seat is prepared to interface with it: this enables the back-seat to subscribe to front-seat sensor streams (e.g., GPS fix, depth, attitude/inclinometer readings) and, conversely, to send positioning or setpoint commands (heading, depth, speed) to the front-seat control loops. Such a separation preserves safety-critical control on the front-seat while allowing flexible, payload-driven development on the back-seat, and it naturally complements the X300’s modular central payload bay and removable computing module.
For the vehicles used only in the experiments described in this paper, the back-seat was implemented on a Raspberry Pi 5 running Ubuntu 20.04 LTS, providing the computational environment required to host the complete back-seat software stack.

The X300 adjusts its buoyancy by taking in water into a dedicated bladder and by controlling the position of the battery pack, which can be shifted forward or aft as required. This dual mechanism enables the vehicle to achieve precise buoyancy regulation and trim control, ensuring optimal stability and maneuverability during underwater operations.

\begin{figure}[h]
       \centering    
       \includegraphics[width=0.80\columnwidth]{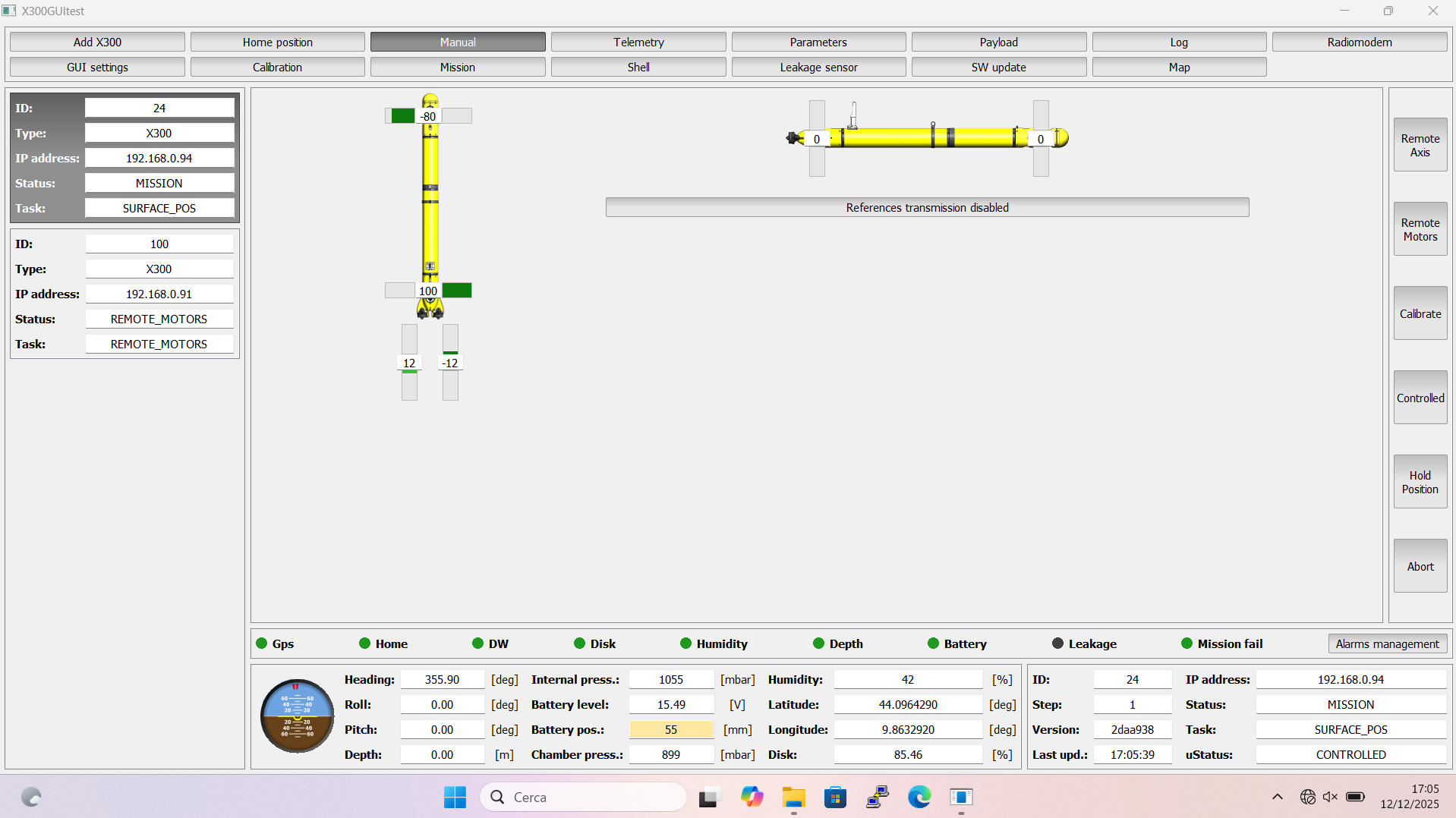}
       \caption{GUI for the operational management of X300\protect~\glspl{auv}. The interface provides tools for mission configuration, real-time monitoring, telemetry visualization, and vehicle control.}
       \label{gui}
\end{figure}

Furthermore, the X300 is equipped with a Wi-Fi antenna, allowing for an efficient communication when the vehicle is on the surface. In this condition, it can be conveniently monitored and controlled through a~\gls{gui} installed on a control station, either located at the dockyard or deployed on board a support vessel. An image of the X300~\gls{gui} is reported in Fig.~\ref{gui}.

The Wi-Fi network can be established either by a surface vehicle or, alternatively, by a portable access point. When a surface vehicle is present within the fleet of vehicles, it typically assumes the role of a gateway, enabling communication both with the shore-based control station and with the AUVs operating underwater. The X300s used in the sea trial are all equipped with a serial medium-frequency acoustic S2C EvoLogics modem, operating in the 18--34\,kHz frequency band. 

\section{System description} \label{systemdescription}

This section presents the experimental system, describing how the interoperability was enabled between~\gls{ros2} and the~\gls{desert} protocol stack (Section \ref{rmw_desert_system_arch}). It also covers the implementation of the acoustic task communication between two nodes (Section \ref{task_implementation}) and the proposed strategy for improving connectivity through environmental sensing and~\gls{auv} mobility (Section \ref{comm_strategy}).

\subsection{System architecture of rmw\_desert} \label{rmw_desert_system_arch}
This subsection introduces the implementation of the~\gls{ros} middleware interface designed for underwater acoustic networks, known as~\gls{rmw_desert}. This publicly available open-source module developed by the Department of Information Engineering at the University of Padova enables~\gls{ros2}-based applications to communicate over the~\gls{desert} Underwater network stack by abstracting and managing key data exchange functionalities at low level, including the node discovery mechanism, the publisher subscriber communication, and the client service paradigm~\cite{RmwDesert}.

\begin{figure}[h]
       \centering    
       \includegraphics[width=0.35\columnwidth]{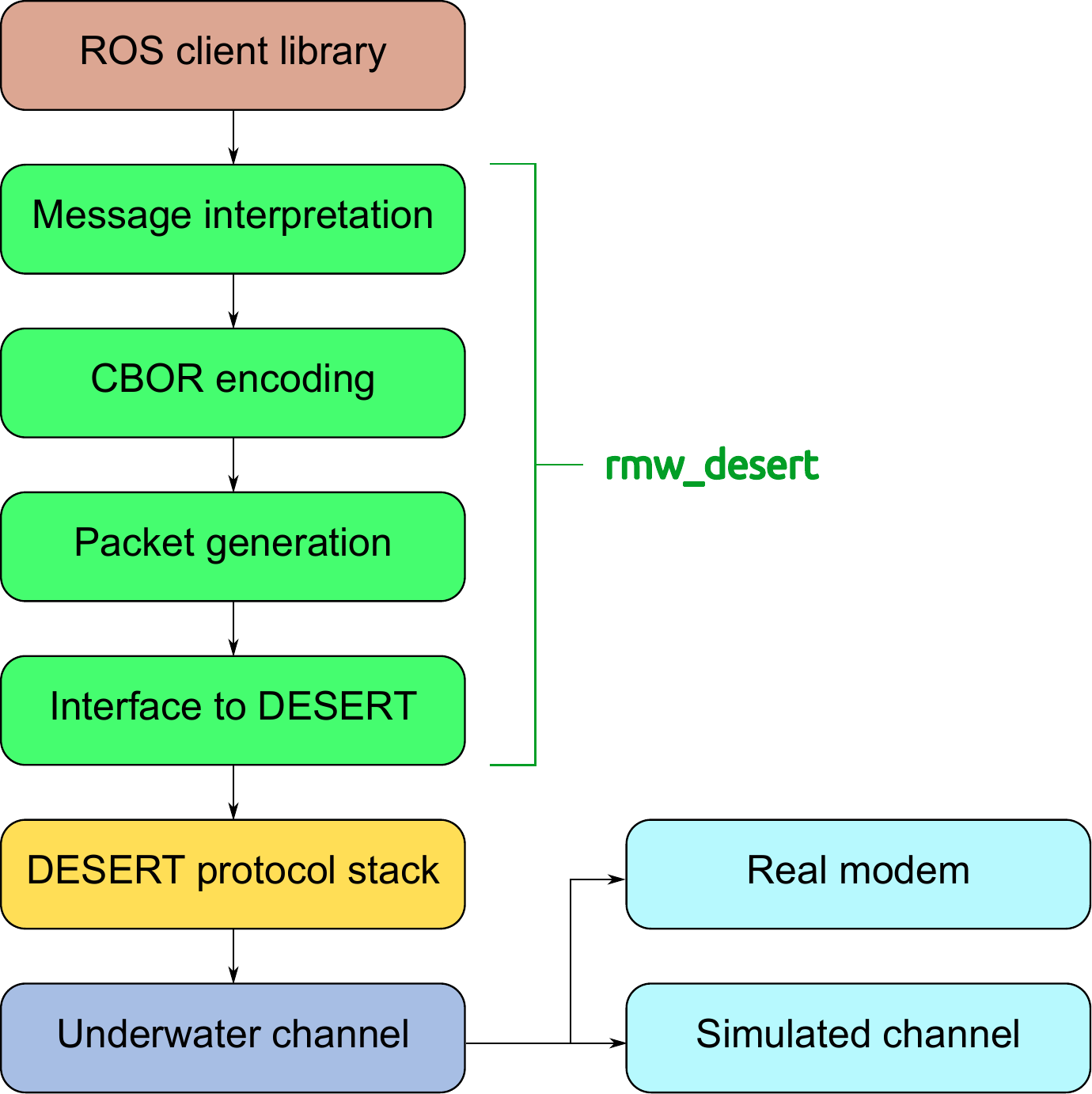}
       \caption{Internal architecture of~\gls{rmw_desert} connecting the ROS client library with the~\gls{desert} framework through the~\gls{ros} middleware interface. The underlying channel can be either a simulation or a physical modem.}
       \label{rmw_desert_arch}
\end{figure}

To understand how these features are supported, it is essential to look at the layered architecture of~\gls{ros2}, focusing on the central interaction point between the~\gls{ros} client library (\verb+rcl+) and the middleware layer, as illustrated in Fig.~\ref{rmw_desert_arch}. When a message is created by a~\gls{ros2} node, the \verb+rcl+ module invokes the middleware interface, passing a pointer to the message data along with a type support handle. This handle provides introspection information that describes the structure, types, and layout of the message to make it possible for the middleware module to correctly interpret all its memory locations.

Using this type information,~\gls{rmw_desert} recursively inspects the memory layout to extract individual fields. These are flattened into a set of primitive data types (e.g., integers, floating-point numbers, booleans, strings), which form the basis for serialization and transmission.

For serialization, the middleware uses~\gls{cbor}, a compact, efficient, and schema-less binary encoding format~\cite{cbor}.~\gls{cbor} is particularly suitable for bandwidth-constrained environments such as underwater communications. It allows complex dynamic data structures to be encoded using minimal space, with fast encoding and decoding performance.~\gls{cbor} uses small headers to describe each data item's type and size. For instance, a small unsigned integer may require only one byte, significantly reducing overhead compared to formats like JSON or XML.

\begin{figure}[!t]
       \centering    
       \includegraphics[width=0.7\columnwidth]{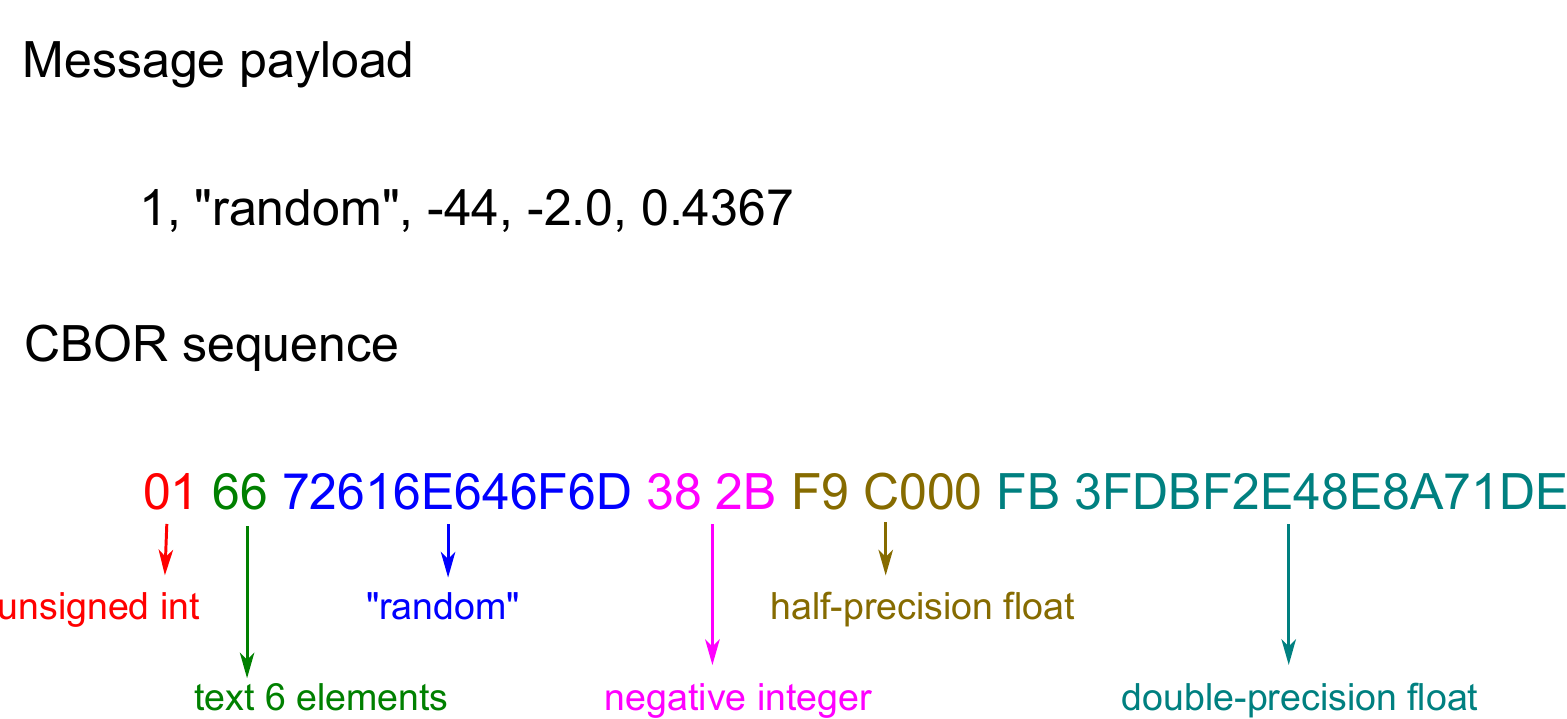}
       \caption{Example CBOR encoding of a message containing five values. Each field is highlighted with a different color, marking the dynamic encoding of the data types based on the value's size.}
       \label{cbor_example}
\end{figure}

An example of message depicted in Fig.~\ref{cbor_example} shows how a series of different data types are serialized in a single stream. Each element is encoded according to its type: a small positive integer is encoded directly in a single byte; a text string is prefixed with a byte indicating the length of the string, followed by its UTF-8 bytes; a small negative integer is represented using some specific marker bits followed by the encoded absolute value; a floating-point number is encoded using the smallest precision that can accurately represent it, such as half-precision for simple values and double-precision for more complex decimals. This structure reflects~\gls{cbor}'s design goal of compact and type-aware serialization.

\begin{figure}[h]
       \centering    
       \includegraphics[width=0.8\columnwidth]{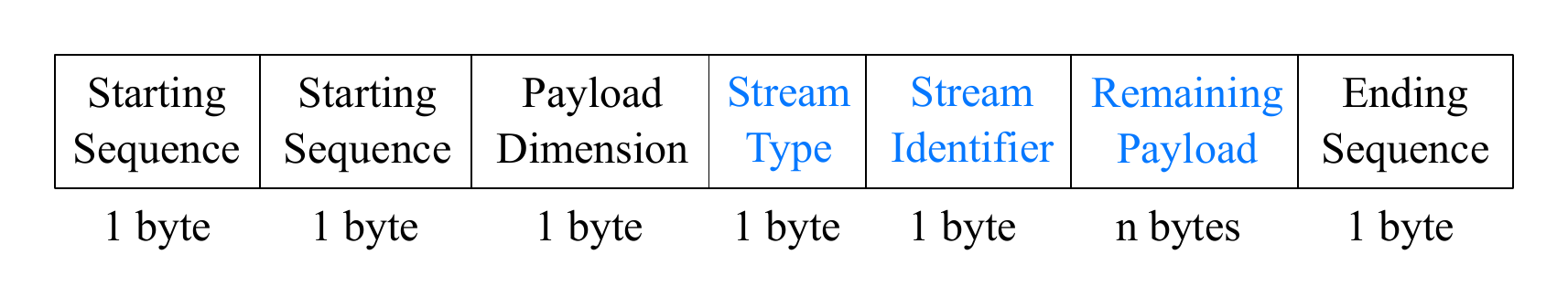}
       \caption{Structure of the data packet used by~\gls{rmw_desert} depicting in blue the CBOR-encoded fields specific for each payload, and in black all the other standard fields required for a correct decapsulation.}
       \label{packet_str}
\end{figure}

Once serialized, the payload is encapsulated in a packet structure that will later be sent to the~\gls{desert} stack, as shown in Fig.~\ref{packet_str}. The packet includes synchronization markers at the beginning and end to facilitate framing in noisy underwater channels. A payload length byte supports integrity checks, while a stream type field identifies the communication role (publisher, service, or client). Additionally, a stream identifier encodes topics or service names in a compact form, allowing each node to route messages appropriately using a static, user-defined configuration file that maps string-based topic or service names to integer numbers, thereby reducing the required field size.

In the final stage, the packet is delivered to the upper layer of the~\gls{desert} protocol stack through a~\gls{tcp} socket managed by the \verb+uwApplication+ module. This module interfaces with the lower layers of~\gls{desert}, which handle~\gls{mac}, routing, and propagation of acoustic signals. The complete stack ensures reliable delivery across challenging underwater environments, characterized by high latency, low bandwidth, and limited connectivity.

\subsection{Task implementation through ROS and DESERT} \label{task_implementation}
\begin{figure}[h]
       \centering    
       \includegraphics[width=0.8\columnwidth]{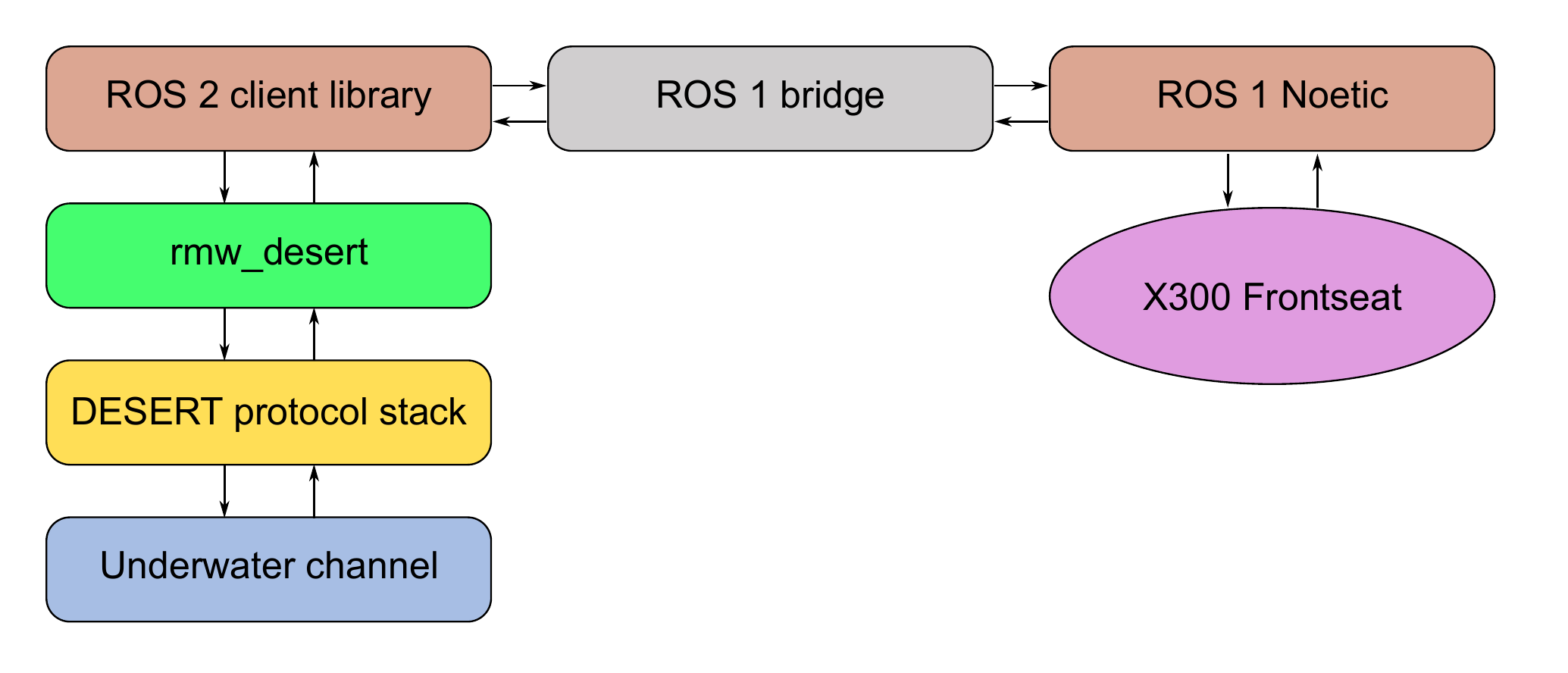}
       \caption{X300 software architecture. The diagram shows the integration of a \glsentryshort{ros2}-based autonomy layer with the \emph{rmw\_desert} middleware and the \glsentryshort{desert} underwater communication protocol stack interfacing with the acoustic channel. A \glsentryshort{ros1} bridge enables interoperability with the legacy ROS~1 Noetic framework running on the X300 front-seat controller, allowing bidirectional exchange of navigation, sensing, and communication data.}
       \label{x300_sw_arch}
\end{figure}

The X300~\glspl{auv} used in this work were delivered with a manufacturer-maintained software environment based on~\gls{ros1} Noetic. This native framework provided stable interfaces for low-level navigation, sensor acquisition, actuator control, and safety-critical functions. However, the development of multi-vehicle mission logic, distributed task allocation, and communication-aware behaviors required capabilities only supported in~\gls{ros2}. To bridge the gap between the vendor-supplied~\gls{ros1} system and the~\gls{ros2}-based communication layer, we adopted a hybrid architecture combining a bridge between the two versions and~\gls{rmw_desert}~\cite{RmwDesert}. The overall architecture and the interaction between its main components are illustrated in Fig.~\ref{x300_sw_arch}.

Each X300~\gls{auv} executed its core operational tasks such as waypoint tracking, depth regulation, and health monitoring within the existing Noetic ecosystem. These tasks relied on well-established topic interfaces and services. Modifying or replacing this stack was not feasible, as it is tightly integrated with proprietary drivers and onboard control loops. Instead, high-level task generation, mission planning, and coordination among the vehicles were implemented in~\gls{ros2}, benefiting from its improved distributed communication and deterministic execution model.

To link the low-level~\gls{ros1} subsystem with the~\gls{ros2} mission layer, we deployed a bidirectional ROS bridge on each vehicle. The bridge exposed selected~\gls{ros1} topics, such as navigation tasks, system diagnostics, and mission progress, to the~\gls{ros2} environment, and conversely translated~\gls{ros2} commands back into~\gls{ros1} message types. Because application-layer communication is strictly point-to-point, every node in the system hosted a dedicated bridge instance that established a direct connection to each of the other nodes. This allowed high-level tasks defined in~\gls{ros2} to be executed by native~\gls{ros1} controllers running in the vehicle's front-seat without modifying their internal logic. Only bandwidth-appropriate necessary data streams were bridged, while other topics (e.g., GPS position) remained local to avoid overwhelming the underwater communication channel. Inter-\gls{auv} task coordination and status exchange were enabled through~\gls{rmw_desert}, in order to incorporate~\gls{desert} Underwater networking functionalities optimized for acoustic modems, including packet scheduling, reliability mechanisms tuned for high-latency links, message bundling, and adaptive transmission policies. By integrating~\gls{desert} at the middleware level, the system retained the standard~\gls{ros} programming model while enabling messages to traverse the acoustic channel efficiently.

Through this arrangement, the mission layer could broadcast task assignments, receive vehicle state updates, and synchronize behaviors across the~\glspl{auv}  despite the constraints of underwater communication. Task-level information like navigation goals, role assignments, and mission-phase transitions were transmitted using~\gls{rmw_desert}, while safety and control loops remained local within each vehicle’s Noetic stack. By exploiting this architecture, all high-level control logic was implemented as a standard~\gls{ros} client application, without embedding any~\gls{desert}-specific functionality. The middleware layer handled all message adaptation, including serialization, compression, packetization, and timing constraints required for the acoustic transmission.

\subsection{Communication strategy for improved connectivity} \label{comm_strategy}
The communication strategy adopted in this work aims to enhance underwater acoustic connectivity by exploiting depth-dependent propagation conditions through real-time environmental sensing. The method extends the environment-adaptive depth-optimization concept we previously investigated in simulation using~\gls{desert} and~\gls{woss}~\cite{Cosimo_Oceans2025}, and translates it into a practical, mission-ready procedure suitable for small AUV teams. The underlying rationale is that, in many underwater scenarios, the ~\gls{ssp} exhibits a pronunced minimum that acts as a favorable propagation layer, often analogous to a weak~\gls{sofar} duct~\cite{Urick}. Operating the vehicles near this depth can reduce transmission loss, mitigate multipath, and ultimately improve packet-level performance.

In the sea trials, the depth-optimization strategy was executed using two cooperating AUVs—a leader equipped with a~\gls{ctd} probe and a follower as the communication partner—while a third platform (i.e., a gateway buoy) was responsible for initiating the optimization sequence. The operational sequence consisted of two main phases. In the first phase, the leader and follower~\gls{auv} operated at randomly selected depths, providing a baseline characterization of link performance under conditions with no environmental awareness. The follower transmitted a burst of one hundred packets of 64 bytes, capturing the behavior of the link when depth is not controlled or adapted.

Once this baseline is established, an external trigger was issued to activate the depth-optimization procedure. Upon receiving the trigger, the leader performed a~\gls{ctd}-based vertical scan while descending from its current operating depth down to a predefined maximum depth. The resulting~\gls{ssp} is analyzed onboard to identify the depth corresponding to the minimum sound speed, which the strategy identifies as the optimal communication depth. Consequently, the leader~\gls{auv} ascends to a shallower staging depth (near-surface) before attempting to distribute the repositioning order. During the subsequent descent toward the optimal depth the leader repeatedly broadcasts the command to move to the \emph{optimal depth} at five seconds intervals, thereby maximizing the likelihood that all followers will receive the instruction. This ascent-and-broadcast procedure is designed to minimize the risk of command loss due to depth-dependent acoustic shadowing. If the leader immediately reaches the \emph{optimal depth} while issuing the repositioning message, a follower located in unfavorable propagation region could fail to receive the instruction and therefore remain out of the~\gls{sofar} duct. Although the adopted procedure may take more time, the probability that all followers successfully detect the positioning order significantly increases.  This mechanism directly addresses the shadowing effects identified in~\cite{Cosimo_Oceans2025}, ensuring a more reliable dissemination of the depth-repositioning command. Once the followers receive the command and converge to the \emph{optimal depth}, they repeat the transmission burst, thereby enabling a controlled and direct comparison between baseline performance and environmentally guided depth alignment.

The proposed strategy offers two main advantages. First, it requires environmental sensing from only one vehicle, reducing the overall sensing burden in a multi-\gls{auv} deployment or manned driven sensing before the operation phase. Second, it relies on lightweight onboard computation, making it feasible on platforms with limited processing capabilities and power consumption restrictions. 

It should be emphasized that this depth-optimization approach is applicable only in specific operational contexts, namely those in which vehicles can afford depth maneuvers to exchange information. For instance, during wreck-search missions or long-term deployment over sufficiently large areas, it can be operationally viable to schedule dedicated optimization windows: in these periods the leader assumes the role of network sink to collect and aggregate data. Conversely, in time-critical missions, in scenarios with tight vehicle dynamics constraints or where continuous connectivity is required, the proposed procedure may be impractical.

\section{Evaluation Setup and Settings} \label{setupandsettings}

This section describes the experimentation evaluation, including preliminary trials conducted to tune the system (Section \ref{preliminary}) and the full-scale sea trial setup used to test the approach in realistic littoral conditions (Section \ref{sea_trial_setup}).

\subsection{Preliminary trials} \label{preliminary}

\begin{figure}[!b]
       \centering    
       \includegraphics[width=0.2\columnwidth]{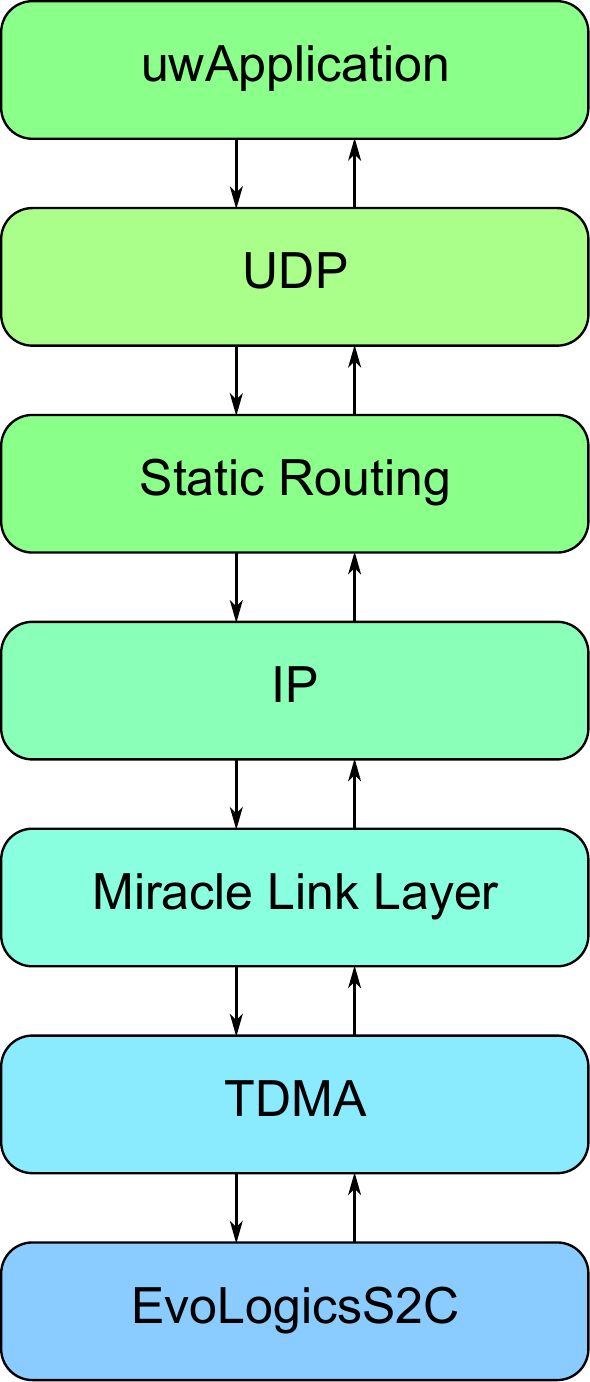}
       \caption{Protocol stack used in the trials. The diagram illustrates the layered communication architecture adopted for underwater networking, spanning from the application layer down to the acoustic modem interface. The entire stack is managed and configured through~\glsentryshort{desert}, which enables modular integration and flexible tuning of protocol parameters to adapt the communication behavior to the underwater channel. }
       \label{protocol_layers}
\end{figure}

In preparation for the initial experimental campaign, we constructed three~\glspl{auv} designed with a fully modular architecture. Each vehicle was assembled from interchangeable components that allowed rapid reconfiguration of its buoyancy system and propulsion modules. This modularity was essential to ensure that the vehicles could reliably transition between positive buoyancy, necessary for surface operations, and neutral or negative buoyancy for submerged navigation during testing. Beyond the core structural elements, each~\gls{auv} was equipped with an EvoLogics acoustic modem to provide underwater communication, enabling real-time monitoring, coordination, and data retrieval during submerged operations. Additionally, a Wi-Fi interface was integrated into the tail section of each vehicle, allowing remote command and configuration while the vehicle was on the surface. This surface-level communication link proved crucial during early tests, allowing rapid adjustments and system checks before each dive. Afterwards the~\gls{desert} Underwater networking framework was integrated within the Raspberry Pi–based modules of the~\glspl{auv}, enabling each vehicle to autonomously manage its full underwater communication stack. This solution allowed us to run the framework on the onboard embedded device, in order to directly manage medium access control, routing, and application-layer protocols, while also interfacing with the mission software. Some improvements were integrated to support the hardware, such as the adaptation for EvoLogics modems with a serial interface. The Raspberry architecture provided sufficient computational resources to run the framework in real time, managing both low-level modem interaction and higher-level coordination tasks.

\begin{figure}[!b]
       \centering    
       \includegraphics[width=0.7\columnwidth]{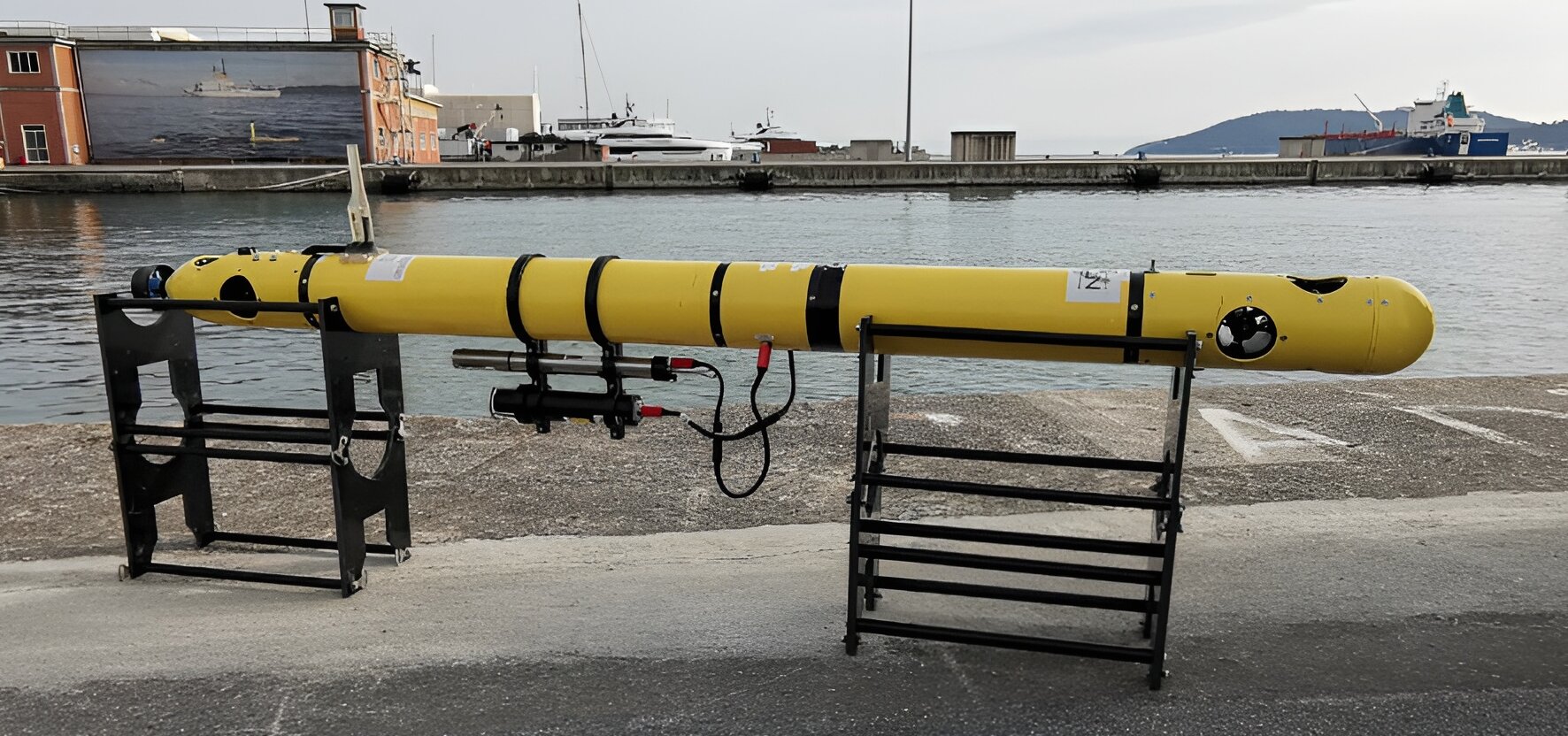}
       \caption{Leader~\protect\gls{auv} used in the experiments, equipped with an Idronaut Ocean Seven 308~\protect\gls{ctd} probe for in-situ environmental sensing and a serial EvoLogics acoustic modem operating in the 18 - 34\,kHz band.
       }
       \label{fig:vehicle}
\end{figure}

To deploy the communication framework across the fleet, dedicated \texttt{tcl} scripts were implemented for each network node, assigning unique identifiers and configuring the corresponding network-layer protocols required by the~\gls{desert} stack as depicted in Fig.~\ref{protocol_layers}. These scripts ensured consistent initialization of the modem drivers,~\gls{mac} and routing modules, and application-level interfaces across all vehicles. At the physical layer, the driver for the Evologics modems was employed, providing a communication link with the device.  At the data link layer, a~\gls{tdma} scheme was implemented, with three distinct slots per frame. This scheme managed access to the shared communication medium, reducing the chances of collision and ensuring fair communication between the vehicles in combination with the NS-Miracle link layer. Then a lightweight IP module tailored specifically for underwater networks was employed, while a static routing protocol simplified the network management by defining fixed paths between nodes, chosen since the network topology was relatively stable. Finally, a simple~\gls{udp} was used at the transport layer.

Furthermore, through~\gls{rmw_desert} a bidirectional link with the~\gls{ros}-based mission software was established, enabling high-level planning and control to interact with the underwater networking processes. With this integration, mission objectives generated within~\gls{ros} could be translated into network-level commands, while feedback from the communication stack was made available to the autonomy layer in real time, supporting coordinated multi-vehicle behavior. Together, these software components and interfaces allowed us to carry out the initial series of tests within the protected bay of the Italian~\gls{cssn} research facility in La Spezia, providing a controlled but operationally representative environment for preliminary trials. After performing the necessary mission-planning adjustments on-site, all components of the system operated without any issues, confirming the robustness of the integrated architecture during this phase.

\subsection{Sea trial setup} \label{sea_trial_setup}
The sea trial was carried out off the Gulf of La Spezia on 19 November 2025 with the objective of testing the depth-optimization strategy in a realistic littoral environment. Experiments were conducted in an offshore area with a seabed depth of approximately 108\,m. Sea conditions at the start of the operations were characterized as Sea State 2~\cite{wmo_manual_codes_2019}; the air temperature at the beginning of the activities was 13\,$^\circ$C. Logistic and safety support was provided by the Italian Navy research vessel Leonardo and a~\gls{rhib}, which ensured continuous monitoring of the~\glspl{auv} throughout each trial.

\begin{figure}[ht]
  
  \centering
  \subfloat[]{\includegraphics[width=0.51\linewidth]{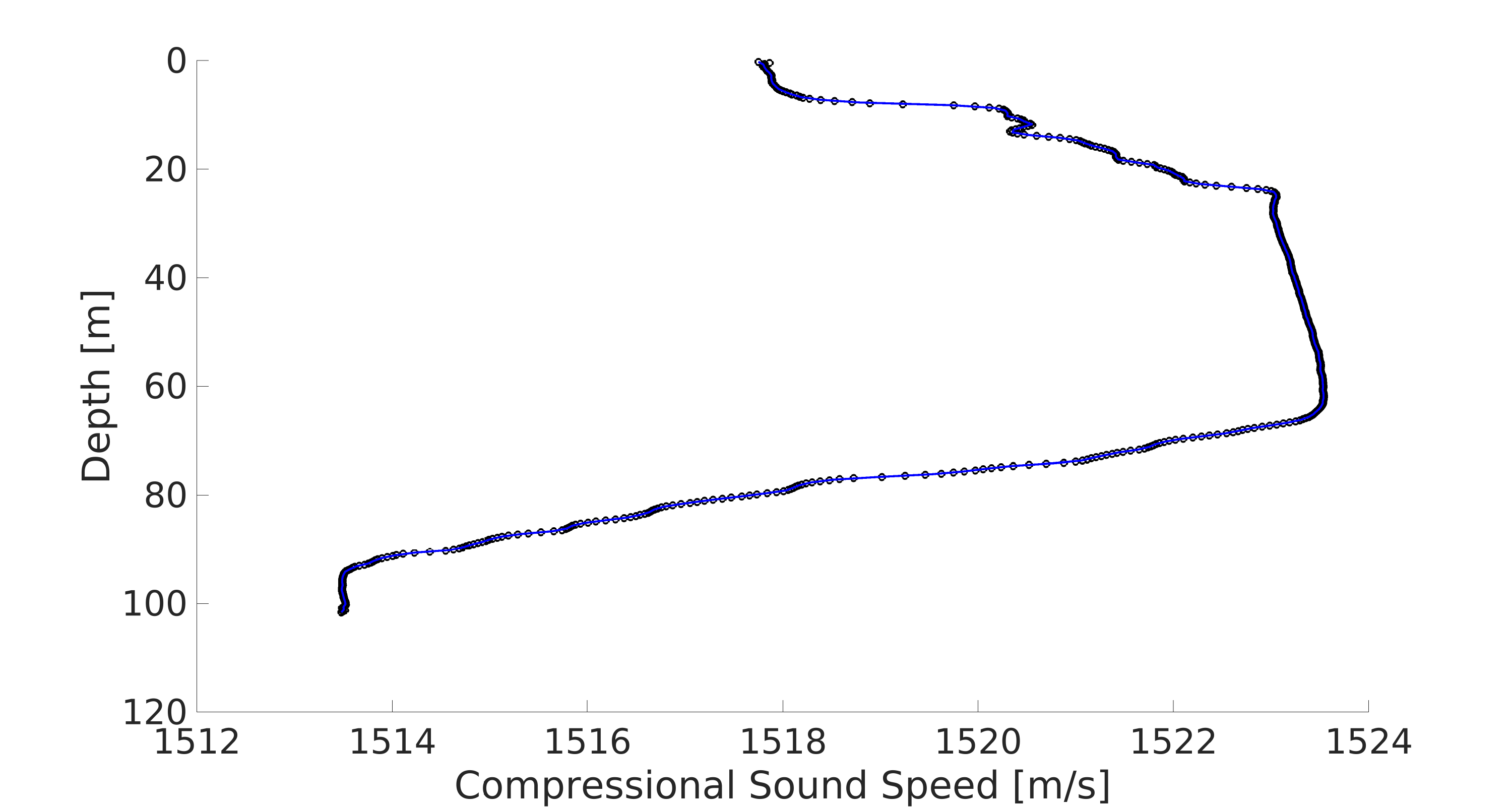}\label{fig:SSP_morning}}
    \subfloat[]{\includegraphics[width=0.51\linewidth]{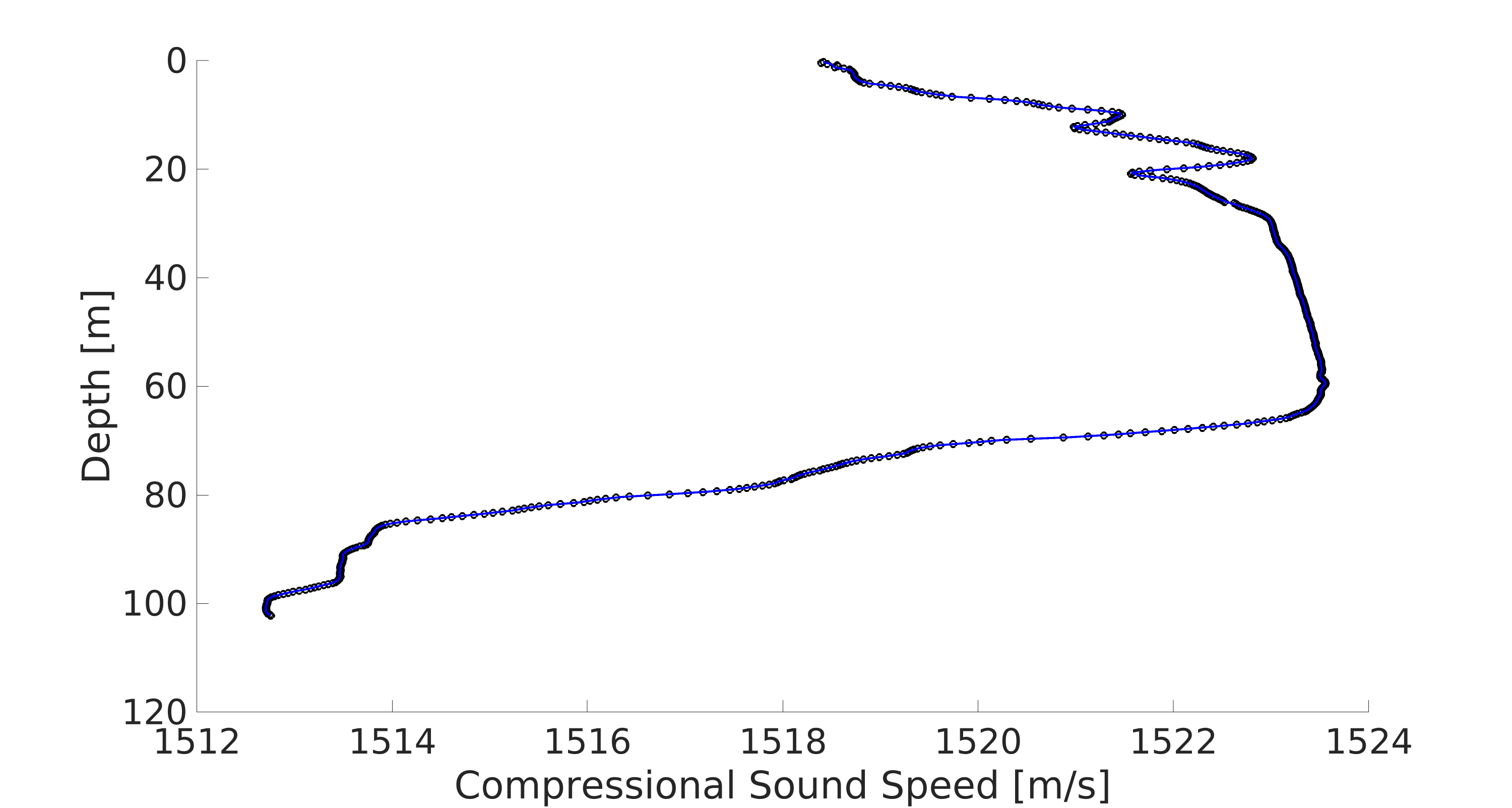}\label{fig:SSP_afternoon}}

    \caption{SSPs measured on November 19 respectively around 9 am~(\ref{fig:SSP_morning}) and 2 pm~(\ref{fig:SSP_afternoon}) in the waters of the Gulf of La Spezia.}
  \label{fig:SSPs}
\end{figure}

\begin{figure}[h]
  \centering
  \includegraphics[width=0.75\linewidth]{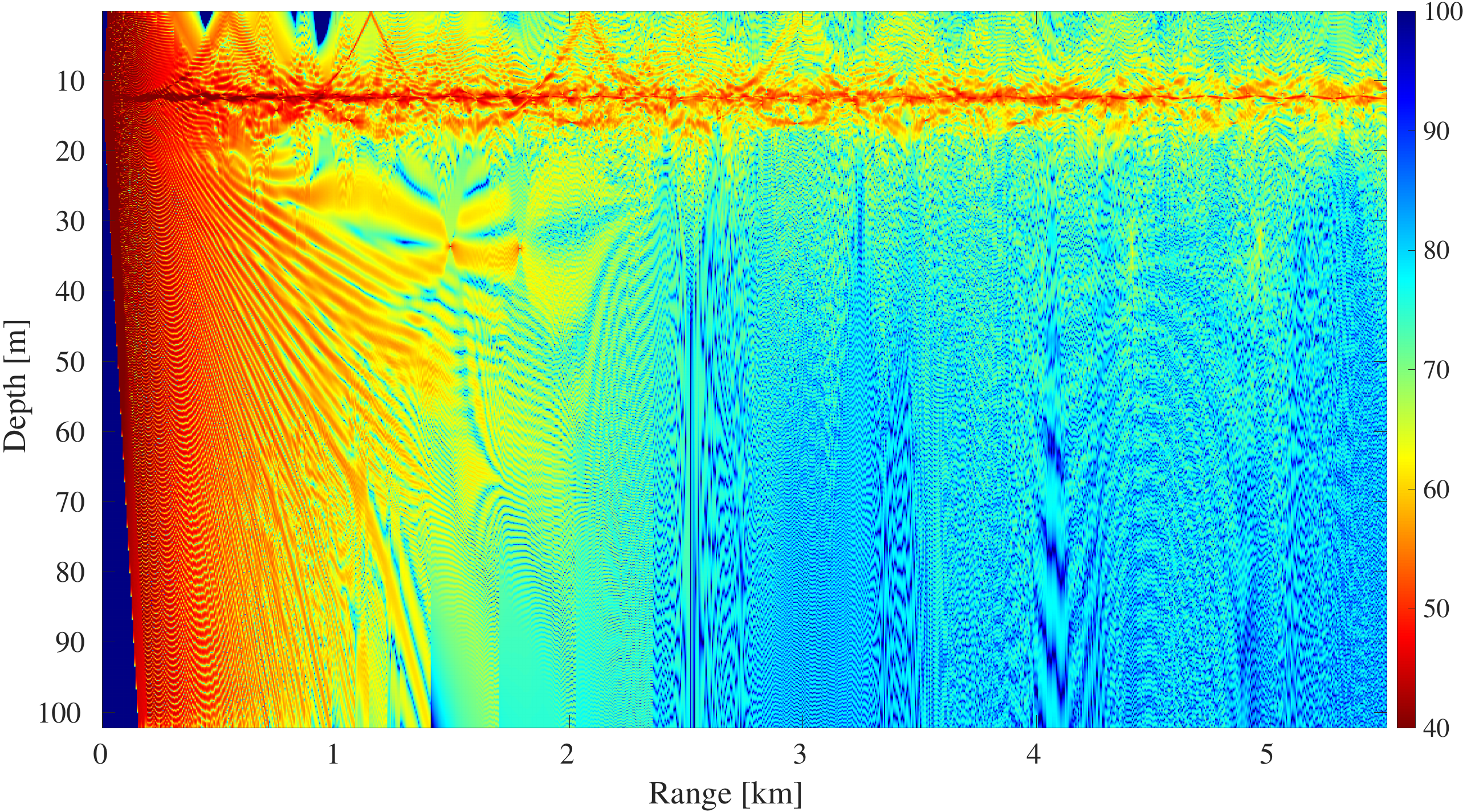}
  \caption{Transmission Loss diagram obtained with Bellhop using the afternoon measured~\protect\gls{ssp}. The source, with a center frequency of 26 kHz matching the EvoLogics modems, is placed at a depth corresponding to the minimum sound speed identified during the tests. The diagram clearly shows the presence of a~\protect\gls{sofar} duct at that depth, guiding acoustic energy over long distances.}
  \label{fig:TransmissionLoss}
\end{figure}

To characterize the vertical acoustic environment,~\glspl{ssp} were collected in two distinct moments of the day using an Idronaut Ocean Seven 310~\gls{ctd}. The corresponding profiles are shown in Fig.~\ref{fig:SSPs}.
Based on the afternoon~\gls{ssp}, the Transmission Loss diagram obtained with Bellhop is reported in Fig.~\ref{fig:TransmissionLoss}, showing that the local minimum in sound speed generates a ducting effect that enhances acoustic energy confinement around that depth.

  


\begin{figure}[!t]
  \centering
    \subfloat[]{\includegraphics[width=0.49\linewidth]{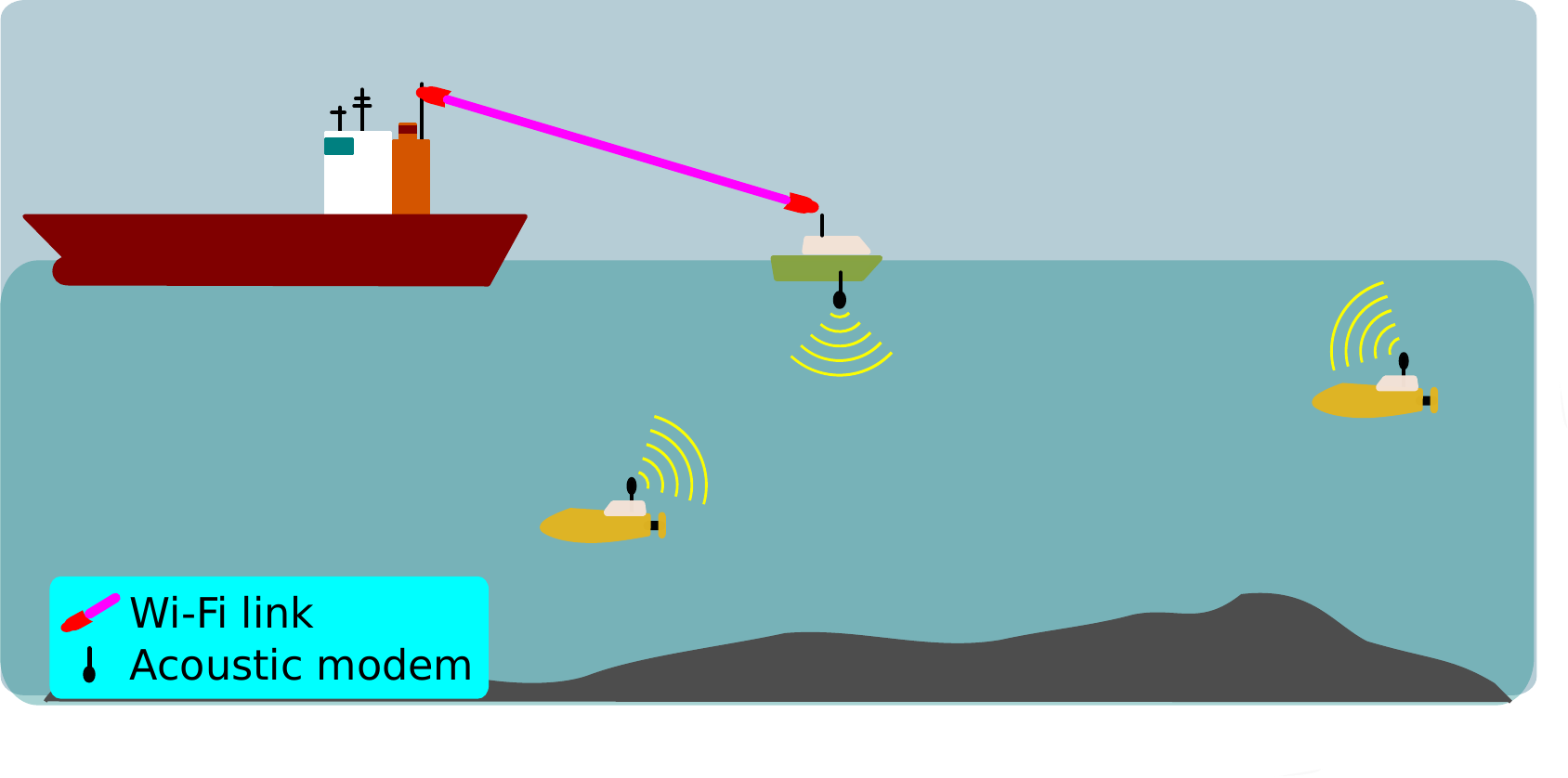}\label{fig:scenario_pre}}
    \subfloat[]{\includegraphics[width=0.49\linewidth]{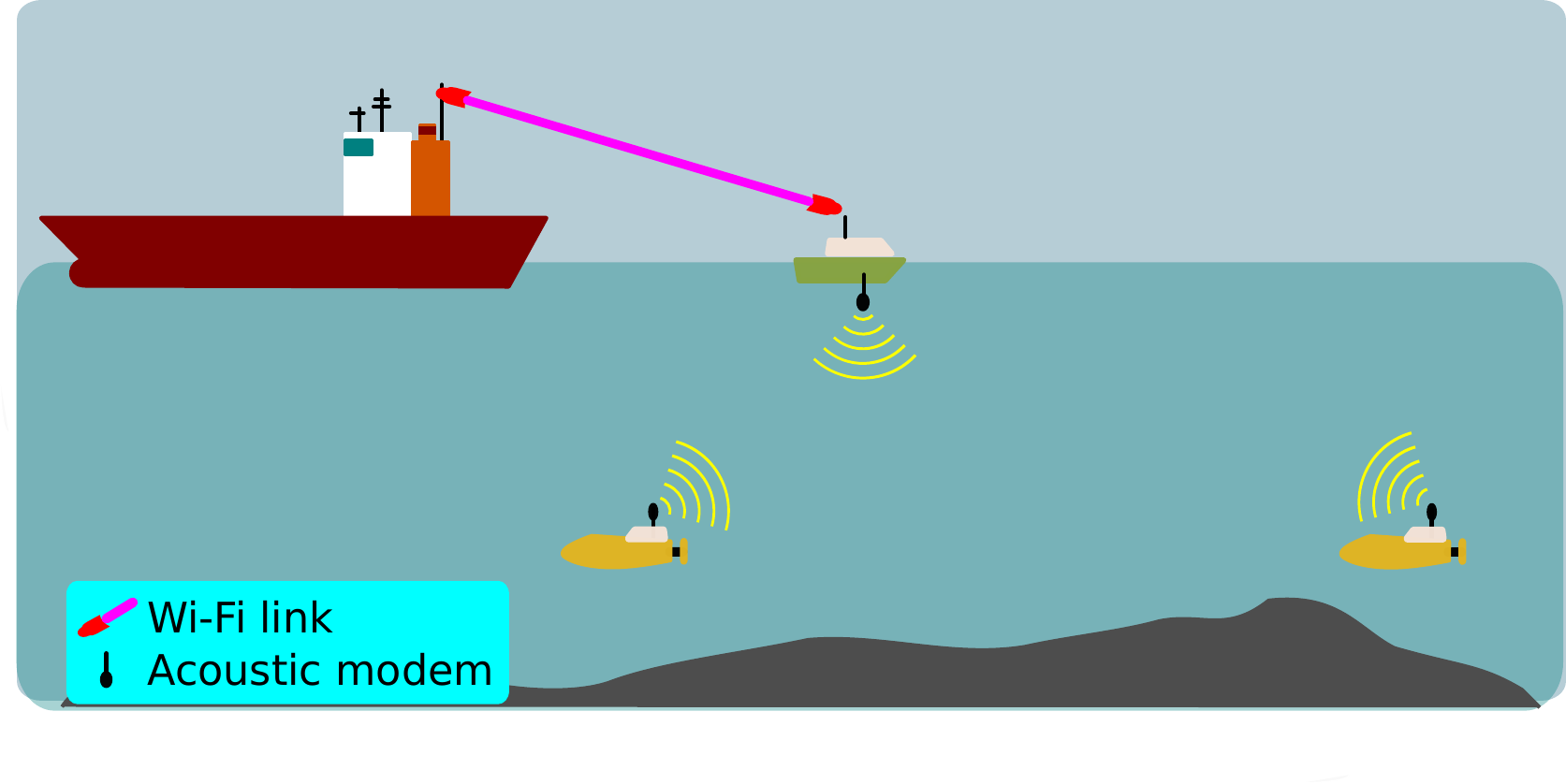}\label{fig:scenario_post}}
    \caption{Scenario of interest, with the surface buoy connected with the ship through a Wi-Fi link and equipped with a modem to command the submerged vehicles. The situation in (\ref{fig:scenario_pre}) represents the initial random-depth condition, while the one in (\ref{fig:scenario_post}) is the optimized one.}
  \label{fig:scenario}
\end{figure}

The trial involved three cooperative platforms as represented in Fig.~\ref{fig:scenario}: a leader~\gls{auv} equipped with an Idronaut Ocean Seven 308~\gls{ctd} probe for in-situ~\gls{ssp} acquisition and for determining the \emph{optimal depth} depicted in Fig.~\ref{fig:vehicle}; a follower~\gls{auv} acting as the communication partner during the tests; and an X300 vehicle configured as a surface buoy and equipped with an EvoLogics acoustic modem for overhearing and trigger delivery. All the nodes are equipped with modems configured with a source level of 187~dB~re~µPa, ensuring sufficient signal strength for the experimental transmission ranges. The buoy maintained a Wi-Fi link with the support vessel, thereby enabling a human-in-the-loop interface: an operator in the vessel's laboratory monitored ongoing transmissions and issued the start of optimization trigger only after the baseline communication burst had terminated  and safe conditions for command dissemination were confirmed.

At the start of each trial, the leader and the follower~\gls{auv} descended to randomly assigned depths. Once stabilized, the follower initiated the baseline transmission sequence, sending 100 packets of 64 \,bytes over a period of approximately eight and a half minutes. Meanwhile, the operator in the support vessel's laboratory—connected via Wi-Fi to the modem on the surface buoy—performed overhearing of the acoustic traffic. After detecting the reception of the hundredth packet, or after waiting for a duration safety exceeding the expected eight and a half minute transmission window, the operator issued a trigger message from the buoy's modem to initiate the network optimization procedure. 

Upon receiving the trigger, the leader~\gls{auv} activated its~\gls{ctd} sensor and performed a controlled vertical descent to a predefined safety depth of 40 \,m to acquire an updated~\gls{ssp}. The vehicle then identified the depth corresponding to the minimum sound speed, adopted as the \emph{optimal depth} for communication. Following the~\gls{ctd} cast, the leader ascended to a near-surface staging depth and, during the subsequent descent to the \emph{optimal depth}, broadcast the repositioning command at five-second intervals. The descent was limited to 40\,m rather than to the seabed to reduce the time required for the test cycle.

After receiving the repositioning instruction, the follower executed the depth adjustment—ignoring any subsequent redundant commands—and transmitted a second burst of 100 packets once stabilized at the \emph{optimal depth}. This allowed a direct comparison between baseline performance at random depths and network behavior under environmentally guided depth alignment.

\section{Results} \label{results}

\begin{table}[ht]
\centering
\small
\caption{Sea trial results including baseline and optimized transmissions. Trial 1 was not completed due to terminal crashes; no data were collected.
For each trial, 100 packets were transmitted at random depths (baseline) and 100 packets were transmitted after convergence to the optimal depth.}
\label{tab:sea_trials_results}
\renewcommand{\arraystretch}{1.2}
\begin{tabular}{c c c c c c c}
\hline
\textbf{Trial} &
\textbf{Inter-node} &
\textbf{Optimal} &
\textbf{Follower depth} &
\textbf{Leader depth} &
\textbf{Packets received} &
\textbf{Packets received} \\
 &
\textbf{distance [m]} &
\textbf{depth [m]} &
\textbf{(baseline) [m]} &
\textbf{(baseline) [m]} &
\textbf{(baseline)} &
\textbf{(optimized)} \\
\hline
1 & 120 & 13.96 & 5 & 12 & 98 & 59 \\
2 & 265 & 13.74 & 5 & 12 & 78 & 35 \\
3 & 1030 & 13.74 & 20 & 12 & 15 & 17 \\
4 & 960 & 15.76 & 24 & 12 & 9 & 15 \\
\hline
\end{tabular}
\label{tab:results}
\end{table}

Table~\ref{tab:results} reports the results of the sea trials, including the horizontal distance between nodes, the \emph{optimal depth} identified by the leader~\gls{auv}, the baseline operating depths of the vehicles, and the number of packets successfully received under both baseline (random-depth) and optimized (depth-adaptive) conditions. For each trial, two transmission phases were performed: 100 packets were transmitted at randomly selected depths and an additional 100 packets were transmitted after convergence to the computed \emph{optimal depth}.

\begin{figure}[h]
       \centering    
       \includegraphics[width=0.75\columnwidth]{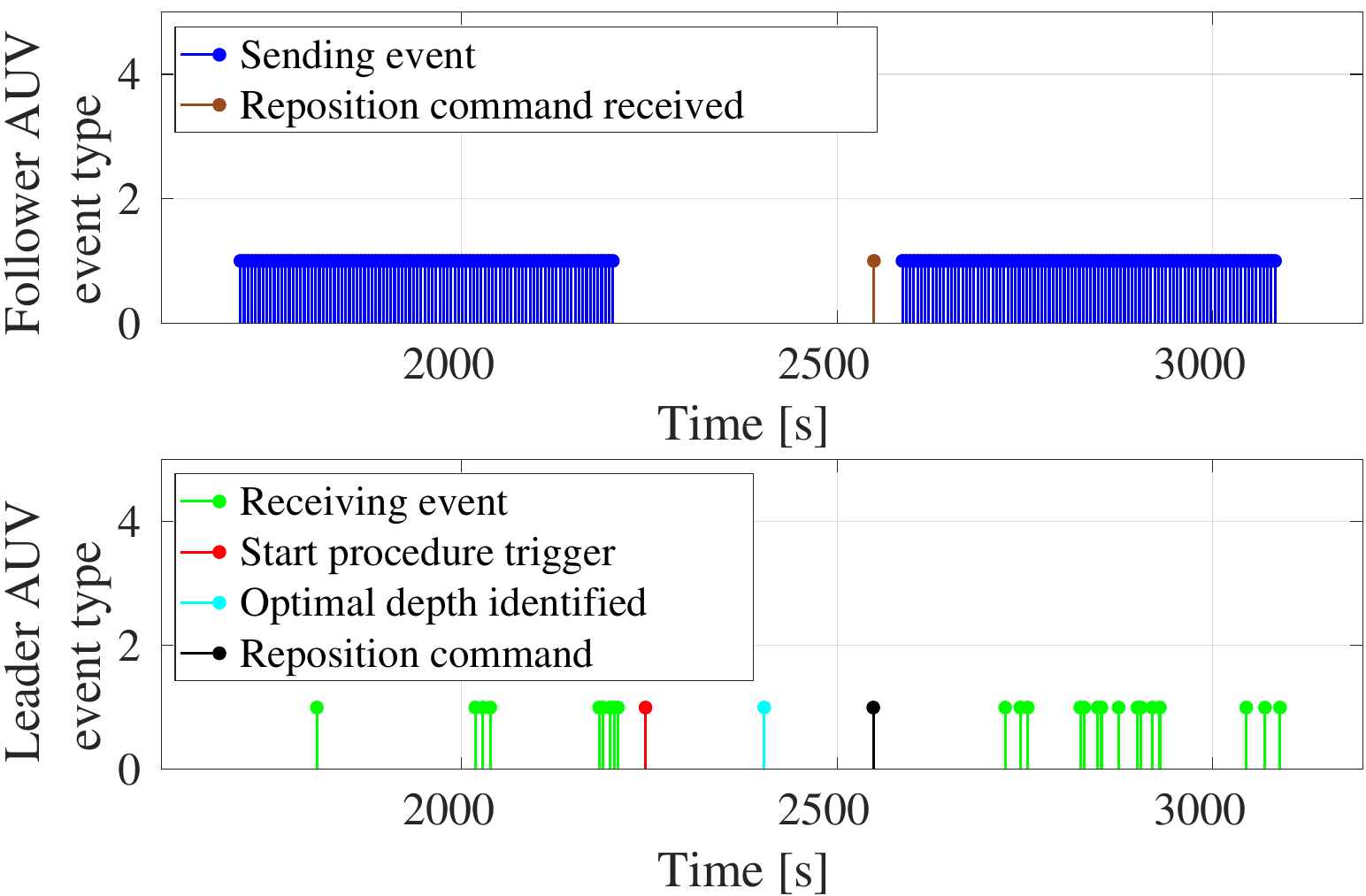}
       \caption{Event timelines derived from~\gls{ros} logs for leader and follower~\gls{auv} during test 4 in Table~\ref{tab:results}, conducted at an initial inter-node horizontal separation of 960\,m. Stem markers indicate packet sending and reception events and key coordination event: reception of the start procedure trigger,~\gls{ctd}-based identification of the \emph{optimal depth}, and transmission of the reposition command. The time gaps between events reflect vehicle maneuvers, including depth adjustment prior to the optimized transmission phase.}
       \label{fig: Results_test5}
\end{figure}

The analysis of the experimental data was performed using logs collected from the~\gls{ros}-based onboard systems. These logs were used to reconstruct the temporal evolution of each trial and to correlate the communication performance with key mission events, such as the start and the end of the transmission phases and the activation of the depth-optimization procedure. In addition, they enabled post-mission alignment of acoustic events with navigation and control states recorded by the vehicles. Fig.~\ref{fig: Results_test5} shows a representative timeline extracted from the~\gls{ros} logs for the sea trial, conducted at an initial inter-node horizontal separation of 960\,m. The figure reports two aligned timelines corresponding to the leader and the follower~\glspl{auv}, and illustrates the main communication and coordination events observed during the experiment.

In the leader timeline, the reception of the start-procedure trigger marks the activation of the depth-optimization sequence. The interval between the trigger reception and the event labeled \emph{Optimal depth identified} corresponds to the~\gls{ctd}-based vertical scan, during which the leader descend to the predefined depth limit to acquire the~\gls{ssp}. The \emph{Optimal depth identified} event indicates the completion of the~\gls{ctd} cast and the identification of the sound-speed minimum. The subsequent delay before the \emph{Reposition command} events reflects the leader's ascent to a shallower staging depth prior to reposition command dissemination.

In the follower timeline, the gap between the reposition command sent by the leader and the onset of the subsequent sending events corresponds to the time required for the follower to adjust its depth and stabilize at the commanded value. By comparing the timestamps of the received packets with the corresponding transmission timestamps recorded in the~\gls{ros} logs, it is possible to estimate the end-to-end packet delay observed during the experiments. This delay accounts for the combined contribution of a small internal processing time within the vehicles—spanning the path from the~\gls{ros} application layer to the physical interface of the acoustic modem—and the acoustic propagation time between the nodes. In the specific case of the trial conducted at an initial inter-node separation of 960\,m, the measured delay was approximately 1.5\,s. The inter-node horizontal distance reported in Table~\ref{tab:results} was estimated from the~\gls{gps} position of the leader and follower~\glspl{auv} at the time of submergence. Once submerged, no absolute positioning was available and no explicit control or compensation for underwater drift was applied. As a consequence, the reported distances should be interpreted as initial separations, with the understanding that unmeasured current-induced drift may have caused variations in the actual inter-vehicle distance during the trials.

An additional source of variability in the observed link performance arose from the orientation of the acoustic transducers, which were mounted in a horizontal configuration (see Fig.~\ref{fig:vehicle}). This arrangement produced a heading dependent gain: the highest received levels were typically measured when the vehicles were aligned stern-to-stern, whereas the lowest levels occurred in bow-to-bow configurations. During the sea trials, no explicit control was exerted over the relative heading or mutual alignment of the vehicles; as a consequence, the variation in transducers orientation contributed to the observed dispersion in communication performance.

\begin{figure}[h]
       \centering    
       \includegraphics[width=0.75\columnwidth]{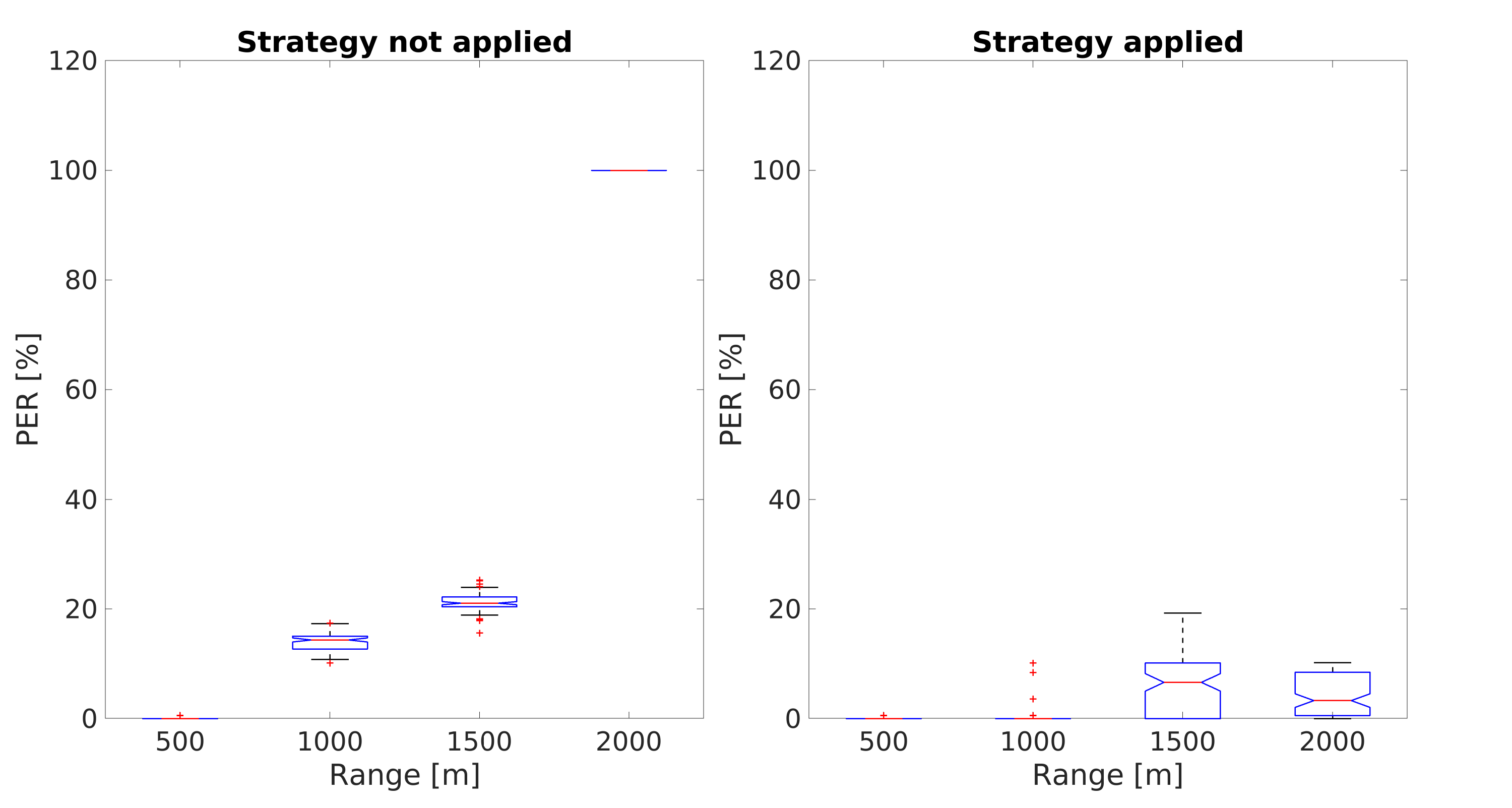}
       \caption{Boxplots of \gls{per} obtained from simulation for baseline (left) and optimized configurations (right) at different inter-node distances (500\,m, 1,000\,m, 1,500\,m, and 2,000\,m). Each box summarizes the distribution over 100 independent runs per scenario, reproducing the environmental conditions measured during the sea trials.}
       \label{fig: PER}
\end{figure}

The results indicate that the proposed depth-optimization strategy begins to provide observable communication improvements when the inter-node separation approaches the kilometer scale. At these ranges, the optimized condition resulted in a higher number of successfully received packets than the baseline case. At shorter inter-node separations, the difference in received signal energy between the baseline and optimized depth configurations is limited; as a consequence, both conditions tend to operate above the demodulation threshold, and no appreciable differences in packet reception are observed. As the range increases, however, small variations in propagation loss translate into quite larger differences in received level and signal-to-noise ratio, making the benefits of depth optimization progressively more evident, as also suggested by the Transmission Loss diagram shown in Fig.~\ref{fig:TransmissionLoss}.

This behavior is quantitatively supported by the simulation study conducted with~\gls{desert}~\cite{Desert, Desert2} integrated with \gls{woss}~\cite{WOSS_Casari,Shallow_zones_SN}, reproducing the environmental condition recorded during the sea trials. Fig.~\ref{fig: PER} reports boxplots of~\gls{per} for different inter-node distances, comparing the baseline and the optimized configurations. Each box summarizes the results for 100 independent runs per scenario.

The results show that the benefits of the proposed strategy become clearly observable at inter-node distances of 1,000\,m and above. At 500\,m the \gls{per} is approximately 0\% even in the baseline condition, indicating that both configurations operate well above the demodulation threshold and therefore exhibit comparable performance. As the distance increases, however, the optimized configuration significantly outperforms the baseline case. At 1,000\,m and 2,000\,m the average~\gls{per} is reduced by more than 95\% with respect to the baseline scenario, while at 1,500\,m the reduction is approximately 74\%. These results confirm that depth adaptive positioning becomes more increasingly beneficial as propagation losses grow with range.

The experimental campaign was constrained to the tested distances by the practical limitations of the setup. Specifically, the Wi-Fi link used for surface supervision restricted the operational envelope, and safety considerations motivated a conservative deployment range. Consequently, while the trials support the effectiveness of depth optimization at kilometer-scale separations, further experiments are required to assess its impact at longer ranges and under different environmental and operational conditions. However, the applicability of the proposed strategy is inherently scenario-dependent. Depth optimization is suitable for missions that allow vertical maneuvers to support data exchange, such as extended-area surveys or operations in which one vehicle acts temporarily as a data sink. Conversely, in time-critical missions or in scenarios with strict depth constraints, the time and energy overhead associated with repositioning may outweigh the communication benefit.

Beyond link-level performance, a key outcome of the sea trials was the successful validation of the complete software architecture. The experiment confirmed the reliability and interoperability of the~\gls{ros}-based autonomy stack, the~\gls{rmw_desert} middleware, the~\gls{desert} communication protocol stack and the software bridges required to interface with the vehicle's legacy~\gls{ros1} framework. The system operated consistently throughout the trials, supporting sensing, onboard processing, communication, and coordinated vehicle actions without requiring ad hoc modifications during field operations. These results demonstrate the feasibility of deploying the proposed strategy on existing~\gls{auv} platforms and highlight the modularity of the adopted software design. The ability to integrate environmental awareness, full control and configurability of the communication protocol stack, and legacy vehicle software through well-defined interfaces provides a solid foundation for future extensions toward increased autonomy and scalable multi-vendor deployment.


\section{Conclusions} \label{conclusions}
This paper presented a real-world validation of the~\gls{ros}–\gls{desert} infrastructure through sea trials involving~\glspl{auv}, demonstrating its effectiveness in realistic and operationally challenging underwater scenarios.
In addition to the experimental validation, the paper also provided a more comprehensive system level description of the\emph{rmw\_desert} middleware and of its integration within the overall architecture, further extending the contribution originally introduced in~\cite{RmwDesert}.
The experimental campaign on the environment-aware depth-optimization strategy conducted in collaboration with the Italian Navy in La Spezia confirmed the correct behavior and integration of all system components, both hardware and software, under real conditions. In particular,~\gls{rmw_desert} proved to be reliable and robust, enabling a transparent interaction between~\gls{ros}-based applications and the~\gls{desert} Underwater networking framework during the trials. The positive outcomes of the sea experiments validate the proposed approach not only from a functional perspective, but also in terms of system-level interoperability and practicality for field deployments. The results show that~\gls{rmw_desert} successfully abstracts the peculiarities of underwater acoustic communications while preserving the~\gls{ros} programming model, allowing developers to exploit existing tools and workflows without resorting to ad hoc solutions tightly bound to specific networking technologies. An additional key advantage of the proposed framework is its ability to support both simulation-based testing and real-world deployments. The same software architecture can be used across simulated and operational environments with minimal reconfiguration, enabling early-stage validation, repeatability of experiments, and risk reduction prior to sea trials. This continuity between simulation and field operation significantly accelerates development cycles and improves the reliability of underwater robotic systems.
  
Furthermore, a major strength of the proposed middleware lies in its dynamic nature:~\gls{rmw_desert} can be loaded at runtime in applications that normally communicate over standard aerial or cabled networks. This capability allows existing~\gls{ros}-based systems to transparently switch to underwater acoustic communications without requiring any modification to the application-level code. As a result, legacy software and well-established robotic software components can be directly reused in underwater scenarios, greatly reducing development effort and promoting software portability. This work highlights how the proposed middleware-based integration can be leveraged to extend the potentialities of~\gls{ros} beyond standard terrestrial networks and beyond environment-specific, custom-designed communication solutions. From an operational perspective, the sea trials also provided the opportunity to assess a concrete, environment-aware communication strategy built on top of the proposed architecture. The tested depth-optimization approach exploits in-situ environmental sensing performed by a single leader vehicle to guide the repositioning of follower~\glspl{auv} toward acoustically favorable layers.

The experimental results primarily demonstrate the successful deployment and operation of the \gls{ros}-\gls{desert} software architecture under real sea conditions, confirming its reliability, interoperability and suitability for a multi-\gls{auv} underwater communication experiment. Within this validated framework, the environmental-aware depth optimization strategy was implemented as a representative use case to assess the potential impact of adaptive vehicle behavior on link performance. The trials indicate that this strategy can yield measurable improvements in packet reception at kilometer-scale inter-node distances, while providing limited benefits at shorter ranges where received signal energy remains sufficiently high. These findings confirm that depth-adaptive communication is not universally applicable, but can be effective in operational scenarios that allow controlled depth variations to support data collection and delivery toward a network sink.

Overall, the presented work shows that the proposed framework integration not only enables robust underwater communication infrastructures, but also provides a practical foundation for deploying adaptive and environment-aware communication strategies in real-world multi-\gls{auv} operations. Since the approach is general and modular, it opens the door to the adoption of~\gls{ros} in other extreme or non-conventional environments, fostering code reuse, scalability, and faster prototyping for future robotic systems operating under severe communication constraints.

\section*{Acknowledgment}
The first two authors contributed equally to this work, investing comparable effort and addressing different and complementary aspects of the research activities. In particular, the first author focused on onboard implementation and experimental activities, including system adaptation and definition of the sea trial, while the second author led the development of ROS-based middleware enabling integration between~\gls{ros2} and~\gls{desert}.

This research was conducted within the framework of SEALab, the joint laboratory between~\gls{cssn} and the~\gls{isme}, which includes the University of Pisa and the University of Padova.

The authors would like to thank the crew of the Italian Navy research vessel Leonardo and the~\gls{cssn} personnel for their support during the sea trial. Special thanks are extended to Mr. Stefano Bianchi, tecnical specialist at~\gls{cssn}, for his close collaboration throughout the entire preliminary testing phase in~\gls{cssn} waters and for his effective support in the hardware setup of the~\glspl{auv}.

\bibliographystyle{IEEEtranDOI}
\bibliography{References.bib}

\end{document}

%% file: acronyms.tex
\newacronym{woss}{WOSS}{World Ocean Simulation System}
\newacronym[plural=AUVs,firstplural=Autonomous Underwater Vehicles (AUVs)]{auv}{AUV}{Autonomous Underwater Vehicle}
\newacronym[plural=UUVs, firstplural=Underwater Unmanned Vehicles (UUVs)]{uuv}{UUV}{Underwater Unmanned Vehicle}
\newacronym{csma}{CSMA}{Carrier Sense Multiple Access}
\newacronym{csmaca}{CSMA/CA}{Carrier Sense Multiple Access with Collision Avoidance}
\newacronym{tdma}{TDMA}{Time Division Multiple Access}
\newacronym{mac}{MAC}{Media Access Control}
\newacronym{per}{PER}{Packet Error Rate}
\newacronym{cdma}{CDMA}{Code Division Multiple Access}
\newacronym{harq}{HARQ}{Hybrid Automatic Repeat Request}
\newacronym{loon}{LOON}{Littoral Ocean Observatory Network}
\newacronym{sofar}{SOFAR}{Sound Fixing and Ranging}
\newacronym[plural=SSPs,firstplural=Sound Speed Profiles (SSPs)]{ssp}{SSP}{Sound Speed Profile}
\newacronym{ctd}{CTD}{Conductivity, Temperature, Depth}
\newacronym{cssn}{CSSN}{Naval Support and Experimentation Centre}
\newacronym{pdr}{PDR}{Packet Delivery Ratio}
\newacronym[plural=UANs,firstplural=Underwater Acoustic Networks (UANs)]{uan}{UAN}{Underwater Acoustic Network}
\newacronym[plural=UASNs,firstplural=Underwater Acoustic Sensor Networks (UASNs)]{uasn}{UASN}{Underwater Acoustic Sensor Network}
\newacronym[plural=IQRs,firstplural=Inter-quartile Ranges (IQRs)]{iqr}{IQR}{Inter-quartile Range}
\newacronym[plural=KPIs,firstplural=Key Performance Indicators (KPIs)]{kpi}{KPI}{Key Performance Indicator}
\newacronym{gui}{GUI}{Graphic User Interface}
\newacronym{ros}{ROS}{Robot Operating System}
\newacronym{ros1}{ROS~1}{Robot Operating System 1}
\newacronym{ros2}{ROS~2}{Robot Operating System 2}
\newacronym{iout}{IoUT}{Internet of Underwater Things}
\newacronym{rf}{RF}{radio-frequency}
\newacronym{mi}{MI}{magnetic induction}
\newacronym{gps}{GPS}{Global Positioning System}
\newacronym{stu}{STU}{Surface Telemetry Unit}
\newacronym[plural={FNOs},firstplural={Fixed Nodes (FNOs)}]{fno}{FNO}{fixed node}
\newacronym{pamp}{PAMP}{Position-Aware Mobility Pattern}
\newacronym{co-pamp}{co-PAMP}{Cooperative-PAMP}
\newacronym{abmac}{AB-MAC}{Adaptive Broadcasting MAC}
\newacronym{moosipv}{MOOS-IvP}{Mission-Oriented Operating Suite–interval Programming}
\newacronym{gram}{GRAM}{Generic Robotic Acoustic Model}
\newacronym{snr}{SNR}{Signal-to-Noise Ratio}
\newacronym{desert}{DESERT}{DEsign, Simulate, Emulate and Realize Test-beds}
\newacronym{rhib}{RHIB}{Rigid-Hulled Inflatable Boat}
\newacronym{rmw_desert}{\texttt{rmw\_desert}}{ROS Middleware for DESERT}
\newacronym{cbor}{CBOR}{Concise Binary Object Representation}
\newacronym{udp}{UDP}{User Datagram Protocol}
\newacronym{tcp}{TCP}{Transmission Control Protocol}
\newacronym{robovaas}{RoboVaaS}{Robotic Vessels as-a-Service}
\newacronym{guwmanet}{GUWMANET}{Gossiping in Underwater Acoustic Mobile Ad-hoc Networks}
\newacronym{guwal}{GUWAL}{Generic Underwater Application Language}
\newacronym{catl}{CATL}{Collaborative Autonomy Tasking Layer}
\newacronym{json}{JSON}{JavaScript Object Notation}
\newacronym{dccl}{DCCL}{Dynamic Compact Control Language}
\newacronym{isme}{ISME}{Interuniversity Center of Integrated Systems for the Marine Environment}